\documentclass[reprint,aps,prd,nofootinbib]{revtex4-1}
\usepackage[utf8]{inputenc}
\usepackage[utf8]{inputenc}
\usepackage{amsmath}
\usepackage{amsfonts}
\usepackage{amssymb}
\usepackage[left=2cm,right=2cm,top=2cm,bottom=2cm]{geometry}
\usepackage{braket}
\usepackage{dsfont}
\usepackage{feynmf}

\DeclareMathOperator{\tr}{\operatorname{tr}}
\DeclareMathOperator{\MPS}{\operatorname{MPS}}
\DeclareMathOperator{\M}{\mathbf{M}}

\DeclareMathOperator{\Tang}{\mathds{T}_{\ket{\psi(A)}}\mathcal{M}_{\MPS}}

\usepackage{graphicx}
\usepackage[bottom]{footmisc}
\usepackage{tikz}
\usetikzlibrary{arrows,decorations.pathmorphing,backgrounds,positioning,fit,petri,shapes,snakes}

\begin{document}

\title{The Kibble Zurek Mechanism of Topological Defect Formation in Quantum Field Theory with Matrix Product States}

\begin{abstract}
The Kibble Zurek mechanism in a relativistic $\phi^{4}$ scalar field theory in $D = (1 + 1)$ is studied using uniform matrix product states. The equal time two point function in momentum space $G_{2}(k)$ is approximated as the system is driven through a quantum phase transition at a variety of different quench rates $\tau_{Q}$. We focus on looking for signatures of topological defect formation in the system and demonstrate the consistency of the picture that the two point function $G_{2}(k)$ displays two characteristic scales, the defect density $n$ and the kink width $d_{K}$. Consequently, $G_{2}(k)$ provides a clear signature for the formation of defects and a well defined measure of the defect density in the system. These results provide a benchmark for the use of tensor networks as powerful non-perturbative non-equilibrium methods for relativistic quantum field theory, providing a promising technique for the future study of high energy physics and cosmology.
\end{abstract}

\author{Edward Gillman}
\email{eg909@ic.ac.uk}
\author{Arttu Rajantie}
\email{a.rajantie@imperial.ac.uk}
\affiliation{Department of Physics, Imperial College London, SW7 2AZ, UK}
%\date{16 October 2017}
\maketitle

\section{Introduction: The Kibble Zurek Mechanism and Tensor Networks}

Classically, \textit{topological defects} are stable, finite energy configurations of a system that interpolate in space between different degenerate choices of vacua (ground state), appearing like ``lumps" of energy with some finite extent (width). In a quantum field theory (QFT), topological defects then appear as particles which are unlike other elementary excitations in that a) they carry a \textit{topological charge} $Q$ associated not to a symmetry of the system, but to the topology of the vacuum manifold and b) they have some characteristic non-zero size \cite{Rajaraman1987,Vachaspati2006}. The existence of a topological charge then separates the theory into different \textit{topological sectors} with the vacuum lying in the $Q=0$ \textit{vacuum sector}. Since topological defects lie outside the vacuum sector, they cannot be studied by standard perturbation theory from within it and in this sense are naturally non-perturbative. However, the equilibrium properties of defects can still be studied semi-classically by starting in the appropriate $Q \neq 0$ sector and non-perturbative equilibrium results can be obtained using lattice methods \cite{Groeneveld1981,Rajantie2010a,Rajantie2012}.

A primary mechanism by which defects are formed is known as the \textit{Kibble-Zurek Mechanism} (KZM) \cite{Kibble1976,Zurek1985} which states that as a system undergoes a symmetry breaking phase transition, if the symmetry broken phase allows topological defects, they will necessarily be formed randomly in the system in a universal manner. The universality of this process comes from the development of a universal length scale $\hat{\xi}$ which is determined in the symmetric phase by the critical exponents of the transition, along with simple dynamical scales. Once in the symmetry broken phase, $\hat{\xi}$ sets the scale of correlated domains between which defects can occur. The defect density is estimated to scale as
\begin{align}
n \approx \hat{\xi}^{- D_{\text{co}}}
\label{DefDen}
\end{align}
where $D_{\text{co}}$ is the co-dimension of the defect, i.e. the difference between the dimension of space and the dimension of the defect (equal to zero for point-like defects, one for line-like defects etc). Since the typical thermalisation time-scale of defects is much longer than other excitations, after some initial relaxation the system will be described by a random distribution of defects. Details of the phase transition that were encoded in
the universal length scale $\hat{\xi}$ will then remain relevant via the defect density $n$ long after the transition has ended. 

The KZM has been confirmed in a number of scenarios theoretically and experimentally, see the reviews \cite{Rajantie2002,DelCampo2014}. While the bulk of this work concerns the KZM during classical (thermal) phase transitions there has been a growing interest in recent years on the KZM during quantum phase transitions \cite{Zurek2005,Dziarmaga2008,Dziarmaga2012,Dziarmaga2015}. In the classical case, the process of topological defect formation is easy to visualise since the configurations of the system can be seen explicitly. Furthermore, the density of defects can be established by explicit counting. However, in a quantum theory, a simple visualisation is not possible and some suitable observable (operator) must be chosen that captures information about defects in the system. While in some simple quantum lattice theories explicit counting operators can be constructed \cite{Dziarmaga2005} counting is generally highly ultra-violet (UV) sensitive and so is inappropriate for use in quantum field theory. Of course, one can still study other quantities such as the correlation length or quasi-particle excitation density during quantum phase transitions in QFT but these do not in general provide information about the density of defects in the system \cite{Uhlmann2010} and, since it is the defects that actually keep the universal physics relevant after the transition has ended, these quantities alone give a somewhat incomplete picture of the KZM.

%Of course, one can still study the development of the universal length scale $\hat{\xi}$ or other quantities such as the quasi-particle excitation density independently of defect formation. However, since it is topological defects that actually keep the universal scale $\hat{\xi}$ relevant following the transition, this is a somewhat incomplete picture of the KZM. In classical lattice systems the number of defects in a system can be explicitly counted, and this can be generalised in some simple quantum lattice theories allowing for a clean calculation of the defect density $n$ \cite{Zurek2005}. However, the situation is more challenging in a quantum field theory where, since it is highly sensitive to the short-distance/ultra-violet (UV) details, explicit counting cannot be used. 

To instead understand the process of defect formation in QFT it was proposed in \cite{Rajantie2010b} that, in the context of a scalar field theory, the equal time two point function of the field in momentum space for a state of random defects will factorise into a contribution coming from the density of defects $n$ and one coming from the width of the defects $d$ which provide the two relevant scales in the system. In the $\phi^{4}$ scalar field theory in $D=(1+1)$ space-time dimensions, which shall be the focus of this paper, the topological defects are known as \textit{kinks}. A system that is driven through a symmetry breaking phase transition should then, following a period of relaxation, be described by a random distribution of kinks and we can assume that the (non-equilibrium) equal time two point function $G_{2}(k) = \braket{\phi(-k)\phi(k)}$ will take the form
\begin{align}
G_{2}(k) = \frac{v^{2}}{n}G_{\text{corr}}(k/n) G_{\text{kink}}(k d_{K}) ~
\end{align}
where $v$ is the vacuum expectation value, $d_{K}$ is the kink width, $G_{\text{kink}}(k d_{K})$ is a factor coming from the spatial profile of the kink and $G_{\text{corr}}(k/n)$ is a factor coming from the spatial distribution of kinks. If this assumption holds, then studying the equal time two point function is particularly attractive since a) the defect density $n$ appears as a long-distance/infrared (IR) observable which is not sensitive to the UV details and b) the finite extent of the defects are clearly manifest through the kink profile term $G_{\text{kink}}(k d_{K})$. This means that the two point function could be used to provide a sensible measure of the defect density in a QFT while also displaying an explicit signature for the formation of defects through the kink profile term.

To confirm the above picture, we would like to study the non-equilibrium time evolution of a quantum field theory as it is driven through a (quantum) phase transition, calculate the two point function and compare it with the expected form. While this can be done explicitly in the classical case, in the case of QFT the essential non-equilibrium calculation of $G_{2}(k)$ is much more challenging and standard non-equilibrium techniques such as the \textit{2-particle irreducible} (2PI) effective action fail to capture the presence of defects \cite{Rajantie2006} requiring more sophisticated variations \cite{Berges2011}. This suggests that a truly non-perturbative, non-equilibrium method is useful and though far less developed than the corresponding equilibrium techniques, such methods do exist e.g. stochastic quantisation for real-time lattices \cite{Berges2005} and Hamiltonian truncation techniques \cite{Rakovszky2016}. 

In this paper, we will focus on the application of \textit{tensor network} (TN) techniques to this problem. This group of techniques allows for the representation of states and observables as tensor networks i.e. as sets of tensors that must be contracted in a particular pattern to obtain the desired result. These representations can then be used to efficiently parametrise a subset of states and observables of physical interest, allowing for the approximation of a wide variety of observables without the need for sampling. Since tensor networks tend to efficiently parametrise certain low entanglement states and can therefore be used to calculate low entanglement approximations to observables, they can be thought of as providing a low entanglement effective theory. While initially specialised to the ground states of gapped (spin) systems with open boundary conditions (OBC) in $D=(1+1)$ via the density matrix renormalisation group (DMRG) algorithm \cite{Schollwock2011}, the scope of tensor network techniques has broadened considerably in recent years. In particular, they have been applied to the study of quantum field theories in $D=(1+1)$ in the lattice regularised setting \cite{Weir2010,Milsted2013,Haegeman2010}, as well using alternative regularisations within a continuous representation \cite{Ganahl2016,Jennings2015,Verstraete2010}. They have also been applied to the study of QFT with gauge symmetries with a focus on the Schwinger model \cite{Banuls2013,Rico2014,Buyens2014,Pichler2016,Buyens2016,Banuls2017},   in addition to some work on other symmetries such as $SU(2)$ \cite{Kuhn2015}. Importantly for our purposes, tensor networks have been further used to study kinks in equilibrium \cite{Haegeman2012,Milsted2013} as well as aspects of the KZM in spin systems \cite{Cincio2007} and the $\phi^{4}$ scalar field theory \cite{Silvi2016}, though without explicitly discussing defect formation.

In this paper, our goals are then twofold. On the one hand we would like to confirm the picture that a QFT undergoing a symmetry breaking phase transition is described by topological defect formation via the KZM and on the other we wish to benchmark the use of tensor network techniques as a non-perturbative non-equilibrium tool for quantum field theory and show that they can capture topological defect formation thus providing a powerful method for studying QFT.

To achieve this, we use the matrix product state (MPS) tensor network, specifically the \textit{uniform matrix product state} (uMPS) to study topological defect formation in the relativistic $\phi^{4}$ scalar field theory in $D = (1+1)$ space-time dimensions. We approximate the time evolution of a state $\ket{\psi}$ using the \textit{time dependent variational principle} (TDVP) technique described in \cite{Haegeman2011,Haegeman2013}. Initially, the state is in approximated in the ground state by within the symmetric phase by a uMPS using the variational-uMPS (VUMPS) algorithm \cite{Zauner-Stauber2017} before being driven though the (quantum) phase transition by a time dependent bare mass $\mu_{0}^{2}(t)$. In the symmetry broken phase, the equal time two point function can then be calculated $G_{2}(k) = \braket{\psi | \phi(t,-k) \phi(t,k)|\psi}$, though we will find it cleaner to instead use the time averaged quantity $\bar{G}_{2}(k)$ with the average taken over some final ``relaxation" period with a time independent bare mass $\mu_{0}^{2}(t_{F})$. We then compare $\bar{G}_{2}(k)$ to the expected form given by a random distribution of defects formed via the KZM, which we call the defect ansatz $G_{\text{def}}(k)$. We find that indeed $\bar{G}_{2}(k) \approx G_{\text{def}}(k)$, under a set of assumptions for the form of $G_{\text{def}}(k)$, and we discuss the nature of these approximations suggesting possible improvements for the future.

The outline of the paper is as follows : In Section \ref {KZM} we review the KZM and explain how it describes the process of topological defect formation, providing details for the $\phi^{4}$ case and describe the form of the defect ansatz $G_{\text{def}}(k)$ which provides  the expected form of the equal time two point function in this case. In Section \ref{TimeEvoMPS} , we review the idea of tensor network representations and how they can be used to calculate low entanglement approximations to a wide set of time dependent observables, essentially providing a low entanglement effective theory. In Section \ref{Res} we then present our results, showing that the obtained form of $\bar{G}_{2}(k)$ is consistent with $G_{\text{def}}(k)$ and the KZM, before concluding in Section \ref{Concl}. 

\section{The KIbble-Zurek Mechanism in the $\phi^{4}$ scalar field theory}
\label{KZM}

\subsection{The Kibble-Zurek Mechanism and Universal Defect Formation}

The equilibrium behaviour of a system undergoing a second order phase transition is well understood. Let us assume that the system is described by a set of dimensionless couplings $\mathbf{g} = \lbrace g_{1},g_{2}...g_{N} \rbrace$ and one of these, e.g. $g = g_{1}$ is being changed by some external process. The relative distance of the system from the critical point can be parametrised by a reduced coupling (reduced temperature) $\epsilon$ such that
\begin{align}
\epsilon = \frac{\left(g - g_{C}\right)}{|g_{C}|} 
\end{align}
and $\epsilon = 0 $ indicates the critical point at $g = g_{C}$. We will consider second order phase transitions that are characterised by the breaking of a global symmetry encoded in the group $G$. In this case the equilibrium state of the system can be characterised by an order parameter $\varphi$ that is zero in the symmetric phase $\epsilon > 0$ where the (unique) equilibrium state is invariant under the symmetry group $G$ but becomes non-zero in the symmetry broken phase $\epsilon < 0$ where there are (degenerate) equilibrium states that are no longer invariant under $G$. In this scenario, the correlation length $\xi$ associated to the order parameter (e.g. by the asymptotic behaviour of the order parameter two point function) diverges such that
\begin{align}
\xi &\approx \xi_{0} |\epsilon|^{-\nu}  ~ , ~ \epsilon > 0 
\label{CE}
\end{align}
near the critical point and the equilibrium state will be characterised in the broken symmetry phase by an infinite correlation length (e.g. by the decay of the order parameter two point function to a constant). 

Let us now consider a scenario where the state of a system is initially in equilibrium in the symmetric phase and evolves under some time dependent reduced coupling $\epsilon(t)$ towards the critical point and into the symmetry broken phase. We then realise that, since no state can have a physical correlation length that increases faster than the speed of light, the state cannot remain in equilibrium all the way to the critical point. This argument, originally made by Kibble \cite{Kibble1976}, leads us to conclude that any physical state of a system, initially in equilibrium, will necessarily become excited when approaching a second order phase transition. Moreover, the actual state must have a finite correlation length when the system is in the broken symmetry phase, in contrast with the equilibrium state of the system.

In general, once in the symmetry broken phase, the state can evolve and we can expect that the correlation length $\xi$ will grow rapidly as the system equilibrates with some thermalisation time-scale $t_{\text{therm}}$. In this way, information about the non-equilibrium dynamics of the phase transition can be wiped out. However, Kibble further argued that in a system whose symmetries allow the presence of topological defects, the state of a system following a symmetry breaking phase transition must contain such excitations. This can again be argued from the requirement of causality. A topological defect corresponds to an excitation which interpolates between different symmetry broken vacua. Therefore, in order to have a state with no defects, symmetry breaking must occur in a spatially uniform manner. In general, this cannot happen since perturbations in causally separated regions must act independently. Additionally, when defects are formed in a system their density $n$ is determined by the correlation length in the system $\xi$ as in Equation (\ref{DefDen}) \cite{Rajantie2002}. The defect density then determines the subsequent correlation length of the state $\xi(n)$ such that the evolution of the correlation length is determined by the evolution of the defects in the system. Often, the time-scales associated to the dynamics of defects e.g. the annihilation of defect-antidefect pairs, are much slower than those naturally associated with thermalisation. As such, information about the dynamics of the phase transition can remain encoded in the distribution of defects well after the transition has occurred and for this reason topological defects are sometimes described as ``fossilised evidence" of the phase transition.

An estimate for the correlation length $\xi$ of the state on entering the symmetry broken phase can be determined by causality. However, a better universal estimate is given by the argument due to Zurek \cite{Zurek1985}. On approach to the second order phase transition, in addition to the diverging length scale, an equilibrium state can have a time-scale $\tau$ characterising the relaxation time which also diverges such that
\begin{align}
\tau \approx \tau_{0} |\epsilon|^{-\mu} ~ , ~ \epsilon > 0 
\label{tauDiv}
\end{align}
near the critical point. If the transition rate is finite and characterised by a (quench) time-scale $\tau_{Q}$ such that $\epsilon = -t/\tau_{Q}$, then an equilibrium state is characterised by both this relaxation time-scale $\tau(\epsilon)$ and a time-scale associated to the rate at which the state is changing due to the time dependence of $\epsilon$. The latter is given by the relative rate of change of $\epsilon$ i.e. $|\dot{\epsilon}/\epsilon|$. In the linear quench case $\epsilon = -t/\tau_{Q}$ we have $|\dot{\epsilon}/\epsilon| = t^{-1}$ so that the associated time-scale is simply the time distance $t$ from the critical point. 

When the relaxation time-scale of the state $\tau$ is much shorter than the time-scale characterising the rate of change of the state, the state will remain in equilibrium (i.e. it evolves adiabatically with the change $\epsilon(t)$). However, since the relaxation time-scale diverges when approaching a critical point, the loss of equilibrium is inevitable and adiabaticity will breakdown at the time $t = \hat{t}$ when
\begin{align}
\hat{t} \approx \tau(\hat{t}) ~ .
\end{align}
If the rate of change $\dot{\epsilon}/\epsilon$ is sufficiently slow, then the breakdown of adiabaticity will occur sufficiently close to the critical point such that $\tau$ can be approximated by its universal behaviour Equation (\ref{tauDiv}). The time $\hat{t}$ at which this breakdown occurs can then be calculated as
\begin{align}
\hat{t} \approx - \left(\tau_{0} \tau_{Q}^{\mu}\right)^{\frac{1}{1+ \mu}} .
\end{align}
Similarly, the correlation length at this time $\hat{\xi}$ can be calculated using Equation (\ref{CE}) to give
\begin{align}
\hat{\xi} &\approx  \xi_{0}  \left(\tau_{0}^{-1} \tau_{Q}\right)^{\frac{\nu}{1+ \mu}} .
\label{KZ_xi}
\end{align}
This length $\hat{\xi}$ is then assumed to equal the correlation length $\xi$ of the state when entering the symmetry broken phase. This assumption is sometimes called the \textit{Adiabatic-Impulse-Adiabatic} (AIA) assumption since $\hat{\xi}$ will be exactly equal to the correlation length in the case that the state ``freezes out" and does not evolve following the time $\hat{t}$ before entering the symmetry broken phase. The development of this universal length scale is sometimes known as the ``Kibble-Zurek Mechanism" in its own right and is important independent of considerations of defect formation in the system. However, as mentioned, defect formation when entering the symmetry broken phase is the primary mechanism by which the universal length scale $\hat{\xi}$ remains relevant after the phase transition has ended : Since $\hat{\xi} \approx \xi$ determines the defect density $n$, which should change only slowly, the universal length scale $\hat{\xi}$ is preserved via the physical correlation length $\xi(n)$ of the state over long periods of time.

\subsection{Universal Defect Formation in the $\phi^{4}$ Scalar Field Theory in $D = (1+1)$}

The KZM as discussed in Section \ref{KZM} can easily be specialised to the $\phi^{4}$ scalar field theory in $D = (1+1)$. The system (theory) can be defined by the action

\begin{align}
S[\phi] = \int dx dt \left[ \tfrac{1}{2}(\partial_{t}\phi)^{2} - \tfrac{1}{2}(\partial_{x}\phi)^{2} - \tfrac{\mu_{0}^{2}}{2} \phi^{2} -\tfrac{\lambda_{0}}{4!} \phi^{4} \right] ~  
\end{align}
which has a single dimensionless bare coupling $g_{0} = \lambda_{0}/\mu_{0}^{2}$ and a global $\mathds{Z}_{2}$ symmetry which acts as $\phi \to -\phi$. The theory exhibits a second-order (quantum) phase transition in the ground state $\ket{\Omega(g)}$. Defining the reduced coupling $\epsilon = \left(g_{0} - g_{C}\right)/|g_{C}|$, then the ground state is unique and $\mathds{Z}_{2}$ invariant in the symmetric phase $\epsilon >0$ but forms a degenerate eigenspace in the symmetry broken phase $\epsilon < 0$ which contains states that break the $\mathds{Z}_{2}$ symmetry.

In the symmetric phase, the characteristic time-scale for relaxation of the ground state is set by the inverse of the gap $\Delta$, which is given by the scalar mass $m_{S}$. Furthermore, Lorentz invariance implies that the gap not only sets the temporal correlation length but also the spatial correlation length such that  $\Delta = m_{S} = \xi^{-1}$. The critical exponents $\mu$ and $\nu$ are then equal and the correlation length $\hat{\xi}$ (\ref{KZ_xi}) is given by
\begin{align}
\hat{\xi} &\approx  \xi_{0}  \left(\Delta_{0} \tau_{Q}\right)^{\frac{\nu}{1+ \nu}}
\label{xiHat_phi4}
\end{align} 
where $\Delta_{0}$ is the coefficient determined by the vanishing gap on approach to the critical point
\begin{align}
\Delta = m_{S} \approx \Delta_{0}|\epsilon|^{\mu} .
\end{align}

Since the phase transition in $D=(1+1)$ is a strong-coupling transition (in the sense that the critical behaviour is not described by a non-interacting theory) mean-field theory cannot be used and breaks down in the vicinity of the critical point. Instead, the critical exponents  are given by the universality class for the $\phi^{4}$ theory which is that of the classical $D = 2$ Ising model. This class has critical exponent $\nu = 1$ such that the state at the breakdown of adiabaticity is characterised by the quantities
\begin{align}
\hat{t} &\approx -\Delta_{0}^{-\frac{1}{2}} \tau_{Q}^{\frac{1}{2}} , \\
\hat{\epsilon} &\approx -\Delta_{0}^{-\frac{1}{2}} \tau_{Q}^{-\frac{1}{2}} , \\
\hat{\xi} &\approx \xi_{0} \Delta_{0}^{\frac{1}{2}}\tau_{Q}^{\frac{1}{2}}  
\label{phi4_xi}
\end{align}
which can be contrasted with the scaling obtained using mean field theory which incorrectly predicts $\nu_{MF} = 1/2$ to give $\hat{t} \sim \tau_{Q}^{1/3}$,  $\hat{\epsilon} \sim \tau_{Q}^{-1/3}$  and $\hat{\xi} \sim \tau_{Q}^{1/3}$.

Under the ``freeze out" or AIA assumption, we will then estimate that the state has a physical correlation length $\xi \approx \hat{\xi}$ when the system enters the symmetry broken phase. In practice, the AIA  does not hold precisely but we can still assume that the scaling of $\hat{\xi}$ will hold so that $\xi \sim \hat{\xi}$. 

In the symmetric phase, the $\phi^{4}$ theory has a single elementary excitation, the scalar particle (i.e. the lowest lying energy eigenstate) with mass $m_{S}$. However, in the symmetry broken phase there are additional particles known as \textit{kinks} that are topological defects with co-dimension $D_{\text{co}} = 1$. In the classical theory, the (anti)kink solutions $\pm\phi_{K}(x)$ to the classical equations of motion are given by
\begin{align}
\phi_{K}(x) = v \tanh \left( \frac{x}{d_{K}} \right) 
\label{kinkConf}
\end{align}
where $v = \sqrt{-6 \mu_{0}^{2} / \lambda_{0}}$ is the classical vacuum expectation value and $d_{K} = \sqrt{ - 2 / \mu_{0}^{2}}$ is the classical kink width. The solutions (\ref{kinkConf}) interpolate between the two classical vacuum solutions $\pm v$ and have a non-trivial topological charge $Q = \pm 1$ which can be calculated from the boundary conditions as
\begin{align}
Q = \frac{1}{2 v} \left( \phi(\infty) - \phi(-\infty) \right) ~ .
\end{align}

When entering the symmetry broken phase we can then estimate the defect density of the state via the KZM and Equation (\ref{DefDen}) to give
\begin{align}
n \sim  \tau_{Q}^{-\frac{1}{2}} . 
\end{align} 

While this result should hold for sufficiently slow quenches, if $\tau_{Q}$ is too small the system will lose equilibrium before ever reaching the critical region and the scaling given by the (quantum) critical exponents will be irrelevant. In this case, mean field theory can be applied and the defect density scales as
\begin{align}
n \sim  \tau_{Q}^{-\frac{1}{3}} ~~ : ~~\tau_{Q} ~ \le ~\tau_{Q}^{X}
\end{align}
where the size of $\tau_{Q}$ at which this quantum-classical crossover takes place $\tau_{Q}^{X}$ can be estimated from the equilibrium data, see \cite{Silvi2016}. 

\subsection{Defect Ansatz}

To study defect formation via the KZM in QFT we need to identify observables that capture information about the defects in the system. In particular, we would like to have an observable that allows for a simple estimate of the defect density to be obtained, while also making the presence of defects manifest, setting them apart from other point like excitations.

Since classically defects in the $\phi^{4}$ scalar field theory correspond to field configurations that interpolate between different sign vacua $\pm v$, one option for extracting the defect density is to simply ``count the zeroes" of the field configurations. However, while this method may be suitable for lattice theories, it is highly ultra-violet (UV) sensitive and instead we would like an observable that allows $n$ to be extracted from long distance data where the UV is irrelevant. 

The equal time two point function $G_{2}(k)$ provides a good observable to study defects. In the classical theory the form of $G_{2}(k)$ for a system of random kinks can be constructed explicitly \cite{Rajantie2006,Rajantie2010b}. The central idea to this construction is that in a system of random kinks there only two relevant scales in the system, $n$ the defect density and $d_{K}$ the kink width. When these scales are well separated ( and typically $d_{K} \ll n^{-1}$ in KZM scenarios) the two point function factorises in momentum space into a contribution coming only from the distribution of kinks $G_{\text{corr}}(k/n)$ and a contribution coming from the kink profile $G_{\text{kink}}(k d_{K})$. The classical two point function for a system of random kinks can then be written as
\begin{align}
G_{\text{RK}}(k) = \frac{v^{2}}{n}G_{\text{corr}}(k/n) G_{\text{kink}}(k d_{K}) ~. 
\label{RandKinks}
\end{align}

In the classical theory the kink profile contribution $G_{\text{kink}}(k d_{K}) = \frac{k^{2}}{4 v^{2}}|\phi_{K}(k)|^{2}$ can be calculated exactly via the Fourier transform of the kink profile $\phi_{K}(x)$ (\ref{kinkConf}) which gives
\begin{align}
\phi_{K}(k) = \frac{2 i v}{k} \frac{\frac{1}{2} \pi k d_{K}}{\sinh{\frac{1}{2} \pi k d_{K}}}  ~
\end{align}
such that 
 \begin{align}
 G_{\text{kink}}(k d_{K}) = \left(\frac{\frac{1}{2} \pi k d_{K}}{\sinh{\frac{1}{2} \pi k d_{K}}}\right)^{2}  ~ . 
 \label{KinkAnsatz}
 \end{align}
 
Additionally, the form of $G_{\text{corr}}(k/n)$ can also be calculated explicitly in the case of uniformly random kinks to give an exponential decay in real space \cite{Rajantie2006}. However, a better form can be found phenomenologically \cite{Rajantie2010b} using classical simulations to give
\begin{align}
G_{\text{corr}}(k/n) = \alpha_{1} e^{- \alpha_{2} (k/n)^{2}} + \frac{\beta_{1}}{[1+\beta_{2}(k/n)^{2}]} 
\label{Def_uniK}
\end{align}
which in real space is just the sum of a Gaussian part and the exponential part coming from uniform randomness
\begin{align}
G_{\text{corr}}(n r) = a_{1} e^{- a_{2} (n r)^{2}} + b_{1} e^{- b_{2} n r} ~ . 
\end{align}

The above picture can be confirmed in a classical field theory by considering the dynamics of an (ensemble) of scalar fields that are driven through a (classical) phase transition before relaxing under some damping term such that the expectation value of the two point function $G_{2}(k) = \braket{\phi(-k)\phi(k)}$ can be calculated and compared to the ansatz for random kinks (\ref{RandKinks}). The ansatz can then be used by first establishing the form of $G_{\text{corr}}(k/n)$. This is achieved by taking a subset of data, explicitly counting the number of defects $n$ and using this to scale the two point function. The assumption $G_{2}(k) = G_{\text{RK}}(k)$ can then be confirmed by rearranging to give
\begin{align}
\frac{n}{v^{2}}\frac{G_{2}(k)}{G_{\text{kink}}(k d_{K})} = G_{\text{corr}}(k/n) ~.
\label{ClassicalFit}
\end{align}
If this holds then the left-hand side of (\ref{ClassicalFit}) should be a universal function of $n$ only and the functional form of $G_{\text{corr}}(k/n)$ can be fit to establish the universal parameters $\alpha_{1},\alpha_{2},\beta_{1},\beta_{2}$. The ansatz (\ref{RandKinks}) can then be used as a one-parameter fit to measure the defect density $n$ in the remaining data, which can then be checked against the values obtained by explicit counting.

When topological defects are formed in a quantum field theory via the KZM, we can again assume that the only two relevant scales in the system are $n$ and $d_{K}$ such that the general factorisation of $G_{2}(k)$ follows as in a classical theory. However, we can expect additional contributions to $G_{2}(k)$ in the quantum theory coming both from the vacuum and the excitations generated during the phase transition : While in a classical theory a damping term can be added to the action to remove energy from the system so that the contribution of excitations can be neglected, in a quantum theory with unitary evolution energy is conserved and we can expect these contributions to be important. 

These additional contributions can then be included to provide a suitable defect ansatz for the case of defects generated by unitary time evolution through a quantum phase transition. Writing the vacuum two point function as $G_{2}^{\Omega}(k)$ and the two point function of the excitations (matter) as $G_{\text{mat}}(k)$ the defect ansatz $G_{\text{def}}(k)$ for a quantum theory can then be written as

\begin{align}
G_{\text{def}}(k) = \frac{v^{2}}{n} G_{\text{corr}}(k/n) G_{\text{kink}}(k d_{K}) + G_{2}^{\Omega}(k) + G_{\text{mat}}(k) ~
\label{DefectAnsatz}
\end{align}
where the various quantities now take on their full quantum corrections. In particular, the ground state $\ket{\Omega}$ determines the vacuum expectation value $v = \braket{\Omega | \phi |\Omega}$ along with the two point function $G_{2}^{\Omega}(k) = \braket{\Omega|\phi(-k)\phi(k)|\Omega}$. Similarly, the one-kink particle state $\ket{K}$ determines the kink profile term $G_{\text{kink}}(k d_{K})$. 

To confirm topological defect formation in the QFT case, we would like to calculate the full non-equilibrium two point function $G_{2}(k)$ and check the assumption that $G_{2}(k) = G_{\text{def}}(k)$ by independently calculating $n , G_{2}^{\Omega}(k) , G_{\text{kink}}(k d_{K})$ and $G_{\text{mat}}(k)$. Similarly to the classical case, this assumption can then be rewritten as

\begin{align}
\frac{n}{v^{2}}\frac{G_{2}(k) - G_{2}^{\Omega}(k) - G_{\text{mat}}(k)}{G_{\text{kink}}(k d_{K})} = G_{\text{corr}}(k/n) 
\end{align}
which should be a universal function of the defect density as before.

However, this not possible in the quantum case in general since the defect density cannot be calculated explicitly by counting and the other quantities such as $v$, $G_{\text{kink}}(k d_{K})$ and $G_{\text{mat}}(k)$ are no longer known exactly.

To overcome the first problem, we will use the assumption $G_{2}(k=0) = G_{\text{def}}(k=0)$ to obtain an estimate of the defect density from the non-equilibrium data. At $k=0$ the contribution from the kink profile drops out of $G_{\text{def}}(k)$ while the matter contributions should be negligible such that $G_{\text{def}}(k=0) \approx v^{2}/n + G_{2}^{\Omega}(k=0)$ and we can define our estimate of the defect density as
\begin{align}
n_{\text{est}} = \left[G_{2}(k=0) - G_{2}^{\Omega}(k=0)\right]/v^{2} ~.
\label{nEst_Def}
\end{align}
This expression then provides a sensible (long distance) estimate of the defect density based on a simple observable, which is highly desirable in its own right. To construct this estimate of the defect density, we then require an approximation of the vacuum expectation values $v$ and $G_{2}^{\Omega}(k)$. These can be calculated relatively easily by e.g. Monte Carlo techniques, though here we will use tensor network techniques for consistency, see Section \ref{Res}. 

With the defect density estimated, we would then like to obtain the form of $G_{\text{kink}}(k d_{K})$. While this can in principle be done using non-perturbative methods, it is more difficult than the corresponding vacuum calculations and in this paper we use a semi-classical approximation by combining the classical kink profile term (\ref{KinkAnsatz}) with the semi-classical width $d_{K} = \sqrt{2}/m_{S}$ where $m_{S}$ is the scalar mass which we can approximate non-perturbatively (see Section \ref{Res}). 

Lastly, we would like to compute the matter term $G_{\text{mat}}(k)$. However, this also is also somewhat difficult to determine and we make further approximations to account for it. In this case we assume that, given sufficient relaxation time, the matter excitations will ``thermalise" in the sense that the two point function $G_{\text{mat}}(k)$ can be approximated by the two point function of a thermal state with the vacuum subtracted i.e.  $G_{\text{mat}}(k) \approx\Delta G_{\text{therm}}(k)$ with $\Delta G_{\text{therm}}(k) = \frac{1}{\mathcal{Z}}\tr\left[\rho \phi(-k)\phi(k)\right] - G_{2}^{\Omega}(k)$ , $\rho = \frac{1}{\mathcal{Z}} e^{- \beta H}$ and $\mathcal{Z} = \text{tr}\left[\rho\right]$. Under this assumption, the matter contributions $G_{\text{mat}}(k)$ are then also given by an equilibrium quantity and there exist non-perturbative methods to evaluate this. However, in the present case we will again use a semi-classical approximation by taking the non-interacting form such that
\begin{align}
G_{\text{mat}}(k) \approx  \frac{1}{\omega_{k}\left(e^{\beta \omega_{k}}-1\right)}
\label{ThermAns}
\end{align}
where $\omega_{k}$ is the non-interacting lattice dispersion relation
\begin{align}
\omega_{k} = \sqrt{\mu^{2} + 4 \sin\left(\frac{p}{2}\right)^{2}} ~
\label{OmegaAns}
\end{align}
and the inverse temperature $\beta$ and mass $\mu$ are treated as free parameters.

Since we do not calculate $G_{\text{mat}}(k)$ a priori and we only know the form of $G_{\text{kink}}(k d_{K})$ approximately, we cannot simply determine the universal part of $G_{2}(k)$ and compare it with $G_{\text{corr}}(k/n)$ as desired. Instead, we will first focus on the region $k/n \ll d_{K}^{-1}/n$ where the contributions from the kink profile and matter should be negligible. Defining the observable
\begin{align}
G_{\text{uni}}(k) &= \frac{n_{\text{est}}}{v^{2}} \left[G_{2}(k) - G_{2}^{\Omega}(k) \right] \nonumber \\
&= \frac{G_{2}(k) - G_{2}^{\Omega}(k)}{G_{2}(k=0) - G_{2}^{\Omega}(k=0) }
\label{GuniQ}
\end{align}
we should then find that $G_{\text{uni}}(k) \approx G_{\text{corr}}(k/n)$ under the assumption that $G_{2}(k) = G_{\text{def}}(k)$ such that the non-equilibrium observable $G_{\text{uni}}(k)$ should be a universal function of $n$ for low $k$ and we can attempt to fit it to the functional form of $G_{\text{corr}}(k/n)$ in this region. We can then compare this to the behaviour of the observable $G_{\text{uni}}(k)/G_{\text{kink}}(k d_{K})$ using the semi-classical approximation of $G_{\text{kink}}(k d_{K})$. If the assumption $G_{2}(k) = G_{\text{def}}(k)$ holds and the semi-classical approximation for $G_{\text{kink}}(k d_{K})$ is accurate, then $G_{\text{uni}}(k)/G_{\text{kink}}(k d_{K})$ should be a universal function of $n$ over a larger region up to $k \approx d_{K}^{-1}$ where we can still neglect the matter term. The fit $G_{\text{corr}}(k/n)$ should then also hold for this larger region and we can use this to estimate the universal parameters $\alpha_{1},\alpha_{2},\beta_{1},\beta_{2}$.

The comparison between the two point function $G_{2}(k)$ and the defect ansatz can then be completed via a two parameter fit using the ansatz for the matter contribution (\ref{ThermAns}) and we should then find that $G_{2}(k) \approx G_{\text{def}}(k)$ over the full range of $k$ and several orders of magnitude in the observable.

\section{Time Evolution with Matrix Product States}
\label{TimeEvoMPS}

\subsection{Tensor Network Representations and Low Entanglement Observables}

It is well known that in general the representation of a quantum state (or observable) is exponentially expensive in the number of degrees of freedom in the system being described. However, the vast majority of states in the Hilbert space are not of physical interest. Rather, the physically important states tend to be highly atypical and are said to form a tiny ``physical corner" of the full Hilbert space. 

For example, the ground states of gapped, local Hamiltonians often have exponentially decaying correlations corresponding in $D=(1+1)$ to the fact that they obey \textit{entanglement area laws} and are in this sense \textit{low entanglement states} \cite{Brandao2015}. Tensor network (TN) techniques leverage this atypicality by providing a representation for states and observables that, while still \textit{complete} so that all states and observables can be represented at exponential cost, are constructed to mimic the real-space quantum correlations (entanglement) of physical states and thus provide an efficient (polynomial cost) representation for this relevant subset of states. 

A simple example of a tensor network is the \textit{matrix product state} (MPS). The MPS can be used to represent the states of lattice systems e.g. a system of $L$ sites with basis $\ket{\mathbf{n}} = \ket{n_{1}}\ket{n_{2}}...\ket{n_{L}}$ where the local Hilbert-space dimension is finite such that $n_{x} = (1,2,...,d)$. The wavefunction $\psi_{\mathbf{n}}$ can then be represented by the nearest-neighbour contraction of $L$ rank-$3$ tensors $M^{n_{x}}_{\alpha_{x},\alpha_{x+1}}(x)$ so that

\begin{align}
\psi_{\mathbf{n_{x}}} &= \sum_{\alpha_{1},\alpha_{2},..,\alpha_{L}} M^{n_{1}}_{\alpha_{1},\alpha_{2}}(1) M^{n_{2}}_{\alpha_{2},\alpha_{3}}(2) ...  M^{n_{L}}_{\alpha_{L},\alpha_{1}}(L) \nonumber \\
& = \tr \left( \M^{n_{1}}(1) \M^{n_{1}}(1) ... \M^{n_{L}}(L) \right) ~ .
\label{MPS_rep}
\end{align}
Denoting the size of the tensors as $(d, \chi , \chi)$ then the uncontracted (external) index corresponds to the local Hilbert-space basis while the contracted (internal) indices of size $\chi$ (often called the ``bond-dimension") correspond to the amount of entanglement in the state. In particular, the states that can be represented by an MPS with bond-dimension $\chi$ have at most an entanglement entropy bound by a constant $S(\rho_{\mathcal{A}}) = \mathcal{O}(\log \chi)$. This means that general states, which have extensive scaling of entanglement entropy, require an exponentially large bond-dimension to represent while one-dimensional entanglement area law states e.g. the ground-state of a gapped local Hamiltonian in $D = (1+1)$, can be represented by an MPS with only polynomial cost, see the review \cite{Schollwock2011} for more information about MPS as well as \cite{Eisert2013} for more details about their entanglement.

Observables can then also be represented as TN. This is typically achieved by contracting together the TN representations for states together with the corresponding TN representations for operators. In particular the \textit{matrix product operator} (MPO) representation is the standard TN representation for operators when using MPS, e.g. see \cite{Schollwock2011}. A simple observable that can be explicitly represented as a tensor network is the state overlap $\braket{\tilde{\psi}|\psi}$.  This  representation can be built by contracting together MPS representations for the states $\ket{\psi}$ and $\ket{\tilde{\psi}}$ such that

\begin{align}
\braket{\tilde{\psi} | \psi} & = \sum_{\mathbf{n_{x}}} \tilde{\psi}_{\mathbf{n_{x}}}^{*} \psi_{\mathbf{n_{x}}} \nonumber \\
& = \sum_{\mathbf{n_{x}}} \tr \left( \prod_{x}  (\tilde{\M}^{n_{x}})^{*}(x) \right) \tr \left( \prod_{x}  \M^{n_{x}}(x) \right) \nonumber \\
&= \sum_{\mathbf{n_{x}}} \tr \left( \prod_{x}  (\tilde{\M}^{n_{x}})^{*}(x) \otimes \M^{n_{x}}(x) \right) \nonumber \\
&= \tr \left( \prod_{x} \left[ \sum_{n_{x}}   (\tilde{\M}^{n_{x}})^{*}(x) \otimes \M^{n_{x}}(x) \right] \right) .
\end{align}
This tensor network has no external indices such that when fully contracted it produces a single number equal to the value of the observable as desired.

Depending on the states in question, different tensor network structures can be chosen such that the appropriate entanglement structure is captured and there are a handful of rigorous results in this regard. For example, the ground-states of one-dimensional gapped lattice systems with local Hamiltonians can be represented efficiently by MPS \cite{Arad2013}. Similarly, higher-dimensional ground-states can be represented by \textit{projected-entangled-pairs-states} (PEPS) while thermal states can be efficiently represented by MPO, assuming an additional bound in the density of states in both cases \cite{Hastings2007}.

While these results show rigorously that tensor networks can be used to represent subsets of the Hilbert space that contain the various states of interest, it is still necessary to actually find those states within this subset. In this regard, there also exist rigorous polynomial-time algorithms in $D = (1+1)$ for finding the MPS representation of the unique ground-state of a local, gapped system \cite{Landau2015} along with polynomial-time algorithms for degenerate ground-states and sub-exponential-time algorithms for low-lying excited states \cite{Arad2017}. However, the power and applicability of tensor network techniques in practice goes well beyond the systems for which rigorous results exist and in general approximations must be used.

Even in cases where an efficient tensor network representation is possible, it is in generally not possible to actually calculate the corresponding observable by fully contracting the tensor network efficiently. Instead, we are limited to the full contraction of only the small subset of observables that can be both efficiently represented and contracted. However, we can still approximate observables using approximate representations and approximate contraction schemes. Since we know it is the low-entanglement states/observables that can be computed exactly with tensor networks we can then view the approximation for generic observables as low-entanglement approximations. Recently, this idea has been formalised by constructing renormalisation group transformations on tensor networks. These methods systematically simplify tensor networks, removing the high-entanglement degrees of freedom and producing a low-entanglement approximation to the observable in question \cite{Evenbly2015,Bal2017,Yang2017}. In this way, we can think of tensor network techniques as providing a low-entanglement effective theory : while high-entanglement observables will be poorly approximated and high-entanglement physics lost, low-entanglement observables and physics can be well approximated, which is precisely the relevant physics in many scenarios of interest. Additionally, these tensor network techniques can be used in real-time and without sampling making for a powerful non-perturbative method.

\subsection{Approximation of Time Dependent Observables with Uniform Matrix Product States}

In this paper, we are chiefly interested in using tensor networks to approximate observables arising from the KZM scenario in the $\phi^{4}$ scalar field theory. In particular, we would like to approximate the equal time two point function 
\begin{align}
G_{2}(k,t) &= \braket{\Omega|\phi(-k,t)\phi(k,t)|\Omega} \nonumber \\
&=\braket{\psi(t)|\phi(-k)\phi(k)|\psi(t)} 
\label{TwoPoint}
\end{align}
 where the time-dependence is generated by the Hamiltonian
\begin{align}
H[\phi,t] &= \int dx \left[ \tfrac{1}{2}\pi^{2} + \tfrac{1}{2}(\partial_{x} \phi )^{2} + \tfrac{\mu_{0}^{2}(t)}{2} \phi^{2} + \tfrac{\lambda_{0}}{4!} \phi^{4} \right] ~  \nonumber \\
\mu_{0}^{2}(t) &= -\frac{t}{\tau_{Q}} + \mu_{0}^{2}(t=0) ~,~ t < t_{F} \nonumber \\
\mu_{0}^{2}(t) &=  \mu_{0}^{2}(t=t_{F}) ~,~ t \ge t_{F} ~ .
\label{HamCont}
\end{align}
The state is initially in the ground-state $\ket{\psi(t=0)} = \ket{\Omega(\mu_{0}^{2}(t=0))}$ and the explicit time-dependence of the Hamiltonian drives the system from the symmetric phase into the symmetry broken phase stopping at $\mu_{0}^{2}(t_{F})$ where the state is allowed to ``relax" by evolving under the final time-independent Hamiltonian.

To approximate the observables of interest, we use the \textit{ uniform matrix product state} (uMPS) tensor network. The uMPS is a special case of the matrix product state tensor network which provides a translationally invariant representation for the states of a lattice system. To apply the uMPS to the QFT we first approximate the QFT by a lattice system with finite dimensional local Hilbert space. This can be achieved by first discretising the continuum Hamiltonian (\ref{HamCont}). Using first-order finite difference approximations for the gradient term, an appropriate lattice theory can be written in lattice units as
\begin{align}
\tilde{H}[\phi] = \sum_{x} \left[ \tfrac{1}{2}(\pi_{x})^{2} + \tfrac{1}{2}(\phi_{x+a} - \phi_{x} )^{2} + \tfrac{\tilde{\mu}_{0}^{2}}{2} \phi_{x}^{2} + \tfrac{\tilde{\lambda}_{0}}{4!} \phi_{x}^{4} \right] 
\label{HamLat}
\end{align}
with $\tilde{H} = a H , \tilde{\mu}_{0}^{2} = a^{2} \mu_{0}^{2} , \tilde{\lambda}_{0} =  a^{2} \lambda_{0} $ and $ \pi_{x} = \partial \mathcal{L}/\partial{(\partial_{t}\phi)} = a (\partial_{t}\phi) $ such that $[\pi_{x},\phi_{y}] = i \delta_{x,y}$. Secondly, the dimension of the local Hilbert space for a site can be truncated by considering only a finite subset of a given basis at a site. This can be done in the field eigenbasis, which allows for rigorous bounds on the associated error \cite{Jordan2014}, but it is computationally more useful to introduce the real-space fock basis via $\phi_{x}= \tfrac{1}{\sqrt{2}} \left( a^{\dag}_{x}+ a_{x}\right)$ and $[a_{x} , a^{\dag}_{y}] = \delta_{x,y}$. The eigenbasis $\lbrace{\ket{n_{x}}\rbrace}$ of the number operator $N_{x} = a^{\dag}_{x}a_{x}$ can then be truncated so that only the subset $\ket{n_{x}} = \lbrace{ \ket{1},\ket{2},...,\ket{d}\rbrace}$ is used and the resulting state space is finite.

Following this truncation, the state $\ket{\psi(t)}$ is approximated as a uMPS. Starting from the MPS representation (\ref{MPS_rep}), a translationally invariant representation can be obtained by requiring that all the rank-$3$ tensors are equal to the same tensor $ M^{n_{x}}_{\alpha_{x} , \beta_{x}}(x) = A^{n}_{\alpha , \beta} $ for all $ x = (1,...,L) $. Then, since the uMPS is defined by a single tensor $A^{n}_{\alpha , \beta} $, it can be used easily for infinite size lattices $L \to \infty$ , see \cite{Haegeman2013} for details.

The uMPS has several computational advantages over the finite lattice MPS. This is mainly due to the fact that, since the boundaries are irrelevant in the $L \to \infty$ limit, an open boundary condition (OBC) uMPS representation can be used freely which offers significant computational advantages over e.g. periodic boundary conditions (PBC). The uMPS representation of a state can then be written as
\begin{align}
\ket{\psi[A]} = \sum_{\mathbf{n_{x}}} v_{L}^{\dag} \left( \prod_{x = -\infty}^{+ \infty} \mathbf{A}^{n_{x}} \right) v_{R} \ket{\mathbf{n_{x}}}
\label{uMPS_def}
\end{align}
where the notation $\ket{\psi[A]}$ emphasises the fact that this state is defined by $ d \chi^{2}$ parameters encoded in the single tensor $A^{n}_{\alpha , \beta} $. The boundary tensors $ v_{L} ,  v_{R}$ are of size $(d,\chi)$ and act as vectors in the matrix product, encoding the (irrelevant) OBC.

The uMPS has been used previously to study the ground state of the $\phi^{4}$ theory and has proved highly successful \cite{Milsted2013}. This is even the case near the critical point where the ground state violates the entanglement area law due to its diverging correlation length. In such situations it is common to use a tensor network representation such as the multiscale entanglement renormalisation ansatz (MERA) that correctly reproduces the logarithmic violation of the area law $S_{\mathcal{A}} \sim \log(L_{\mathcal{A}})$ \cite{Vidal2006}. However, the uMPS was shown to correctly capture the contributions of kink-antikink excitations to the ground state observables in the vicinity of the critical point, a fact which suggests its potential for use in the KZM of defect formation \cite{Gillman2017}.

To approximate the state $\ket{\psi(t)}$ with a uMPS, the ground state is first approximated. This can be achieved by minimising the energy of the uMPS by varying its $d \chi^{2}$ degrees of freedom encoded in the tensor $A^{n}_{\alpha , \beta}$ e.g. by adapting standard techniques such as the conjugate gradient algorithm \cite{Milsted2013}. However, in this paper we use the highly efficient variational-uniform-matrix-product-state (VUMPS) algorithm \cite{Zauner-Stauber2017}  which mimics more standard MPS variational energy minimisation techniques.

The time dependent state $\ket{\psi(t)}$ can then be approximated by evolving the initial ground state uMPS approximation. Of course, one would like to carry out the full time evolution corresponding to the application of the unitary time evolution operator or the solution to the Schr\"{o}dinger equation
\begin{align}
\frac{d}{dt} \ket{\psi(t)} &= - i H \ket{\psi (t)} \nonumber \\
\ket{\psi(t=0)} &= \ket{\Omega[A]} ~ . 
\label{SE}
\end{align}
However, even if we begin from a uMPS with bond-dimension $\chi$, in general the state obtained by such an evolution will no longer be a uMPS of size $\chi$ but rather described by some larger bond-dimension $\chi'$. Physically, this corresponds to the fact that time evolution tends to increase the entanglement in a state, as found when studying the time evolution of states under sudden ``quenches" of the Hamiltonian \cite{Calabrese2005}. In the sudden quench case, the entropy of entanglement can increase linearly with time $ S_{\mathcal{A}} \sim t$ which would require a uMPS of size $\log(\chi) \sim t$ to represent exactly i.e. one with exponentially increasing bond-dimension. Therefore, to approximate the time evolution, one must consider only an evolution within the subset of states represented by uMPS with bond-dimension $\chi$. This approximation then throws out the high entanglement degrees of freedom and we can expect certain high-entanglement observables to be poorly approximated while still approximating the low-entanglement observables. 

With MPS, an approximation of time evolution can be achieved by first breaking up the unitary time-evolution operator into small time-steps and applying one operator at a time, allowing the bond-dimension to grow $\chi \to \chi'$, before truncating the MPS back down to the starting subset $\chi' \to \chi$. This idea is quite intuitive and implemented in its most established form by the ``time evolving block decimation" (TEBD) algorithm \cite{Vidal2003}, though a number of other related methods exist, see e.g. \cite{Wall2012}.

Another approximation of the time evolution can be achieved by realising that the subset of states defined by the MPS in a number of cases (including the infinite size uMPS case) forms a smooth manifold $\mathcal{M}_{\MPS}$ \cite{Haegeman2014a}. As such, in order for a time evolution to stay within this subset of states, only tangent vectors $\ket{\Phi} \in \Tang$ to the current state $\ket{\psi[A]}$ can be used to update the state. Thus, the full time-evolution of the state can be approximated by projecting the right-hand-side of the Schr\"{o}dinger equation (\ref{SE}) down to the tangent space of the state. The projector to the tangent state at this point can be written as $\hat{P}_{\mathds{T}_{\ket{\psi(A)}}\mathcal{M}_{\MPS}}$ such that the desired evolution is given by the equation
\begin{align}
\frac{d}{dt} \ket{\psi(t)} &= - i \hat{P}_{\mathds{T}_{\ket{\psi(A)}}\mathcal{M}_{\MPS}}\left[\hat{H} \ket{\psi (t)}\right] ~ .
\label{SE_UMPS}
\end{align}
Such a projection is equivalent to finding the tangent vector (state) $\ket{\Phi} \in \Tang $ which satisfies the minimisation problem
\begin{align}
\min_{\ket{\Phi}} || \ket{\Phi} + i \hat{H} \ket{\psi (t)} ||^{2} .
\label{Min_tang}
\end{align}
This can be solved explicitly by finding representations for the tangent vector $\ket{\Phi}$ as a tensor network. Since the tangent space is spanned by the set of $ d \chi^{2}$ partial derivatives $\frac{\partial}{\partial A_{\alpha,\beta}^{n}}\ket{\psi} = \ket{\partial_{i} \psi}$, a tangent state can be written as a sum of these basis elements
\begin{align}
\ket{\Phi[B]} = B^{i}\ket{\partial_{i} \psi}
\label{TangVec1}
\end{align}
and is therefore specified by the coefficient tensor $B^{i} = B^{n}_{\alpha,\beta}$ of size $d \chi^{2}$. Since $\ket{\psi}$ is defined by a uMPS (or MPS) the partial derivative can then be evaluated such that the tangent vector can then be written as a sum of states $\ket{\Xi(m)}$ which are equal to $\ket{\psi[A]}$ except that the tensor at site $x=m$ is replaced by the coefficient tensor $B^{n}_{\alpha,\beta}$ and
\begin{align}
\Xi_{\mathbf{n_{x}}}(m) = \left(\prod_{x = - \infty}^{m-1}\mathbf{A}^{n_{x}}\right) \mathbf{B}^{n_{m}}\left(\prod_{x = m+1}^{\infty}\mathbf{A}^{n_{x}}\right) ~ .
\end{align}
The tangent vector is then given by
\begin{align}
\ket{\Phi[B]} = \sum_{m} \ket{\Xi(m)} ~.
\label{TangVec2}
\end{align}
In fact the basis of partial derivatives is overcomplete corresponding to the well known gauge freedom in MPS and must be restricted to achieve a useful implementation, see \cite{Haegeman2014b} for details.

The representation of the tangent vectors (\ref{TangVec2}) allows for an explicit expression of the tangent space projector $\hat{P}_{\mathds{T}_{\ket{\psi(A)}}\mathcal{M}_{\MPS}}$ to be found. The evolution equation (\ref{SE_UMPS}) can then be written in terms of the update of the tensor $\mathbf{A}^{n}$ alone to give
\begin{align}
\dot{\mathbf{A}}^{n} = - i \mathbf{\tilde{B}}^{n} ~ 
\label{SE_TenUpd}
\end{align}
where $\mathbf{\tilde{B}}^{n}$ is constructed such that $\ket{\Phi[\mathbf{\tilde{B}}^{n}]}$ satisfies the minimisation problem (\ref{Min_tang}).

\begin{figure*}[t]
\centering
\includegraphics[width=0.8\linewidth]{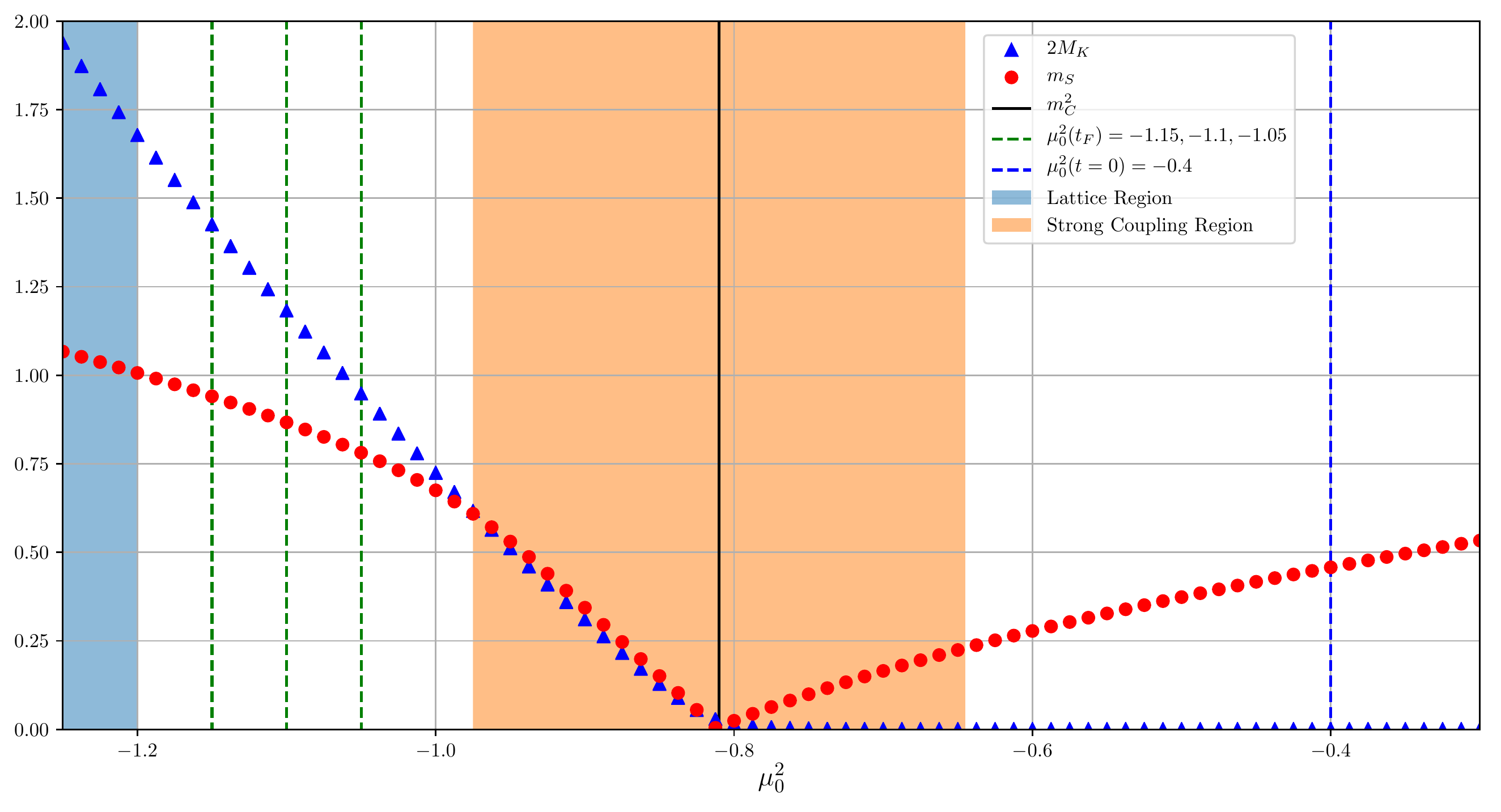}
\caption{The scalar mass $m_{S}$ (red circles) and twice the kink mass $2 M_{K}$ (blue triangles) as estimated using the tensor network techniques described in \cite{Vanderstraeten2015} with $d=18,\chi=16$ and \cite{Gillman2017} with $d=18,\chi=14,L=32$ respectively. These quantities map out the important parameter regions studied by sweeping $\mu_{0}^{2}$ for a fixed $\lambda_{0} = 3$. The leftmost shaded region corresponds to the ``lattice region" where $m_{S} >1 $ such that lattice effects are important and should be excluded to get a good comparison with the KZM. Furthermore, the initial and final bare masses $\mu_{0}^{2}(t=0)$ and $\mu_{0}^{2}(t_{F})$, indicated by the dashed vertical lines, should lie outside the shaded ``strong-coupling region" where, in the broken symmetry phase, $m_{S} \approx 2 M_{K}$ and the kink-antikink excitations behave as standard scalar excitations.
}
\label{EquilMS}
\end{figure*}

While initially this time evolution procedure looks quite different to the more familiar MPS time evolution procedures, it is in fact closely related as shown in \cite{Haegeman2014b}. It is also quite attractive since the optimal truncation of bond-dimension with the time-evolution is encoded automatically in the first-order (highly non-linear) differential equation (\ref{SE_TenUpd}).

The time dependent state $\ket{\psi(t)}$ can now be approximated by first finding an approximation to the ground state as a uMPS and then updating the state according to the equation (\ref{SE_TenUpd}) which we achieve using a $5^{\text{th}}$ order Runge-Kutta scheme. We note that the above ``geometric picture" of the time evolution approximation can also be derived using the ``time dependent variational principle" (TDVP) \cite{Haegeman2011} which, while perhaps less intuitive, can be more widely applied.

\begin{figure*}[t]
\centering
\includegraphics[width=1.0\linewidth]{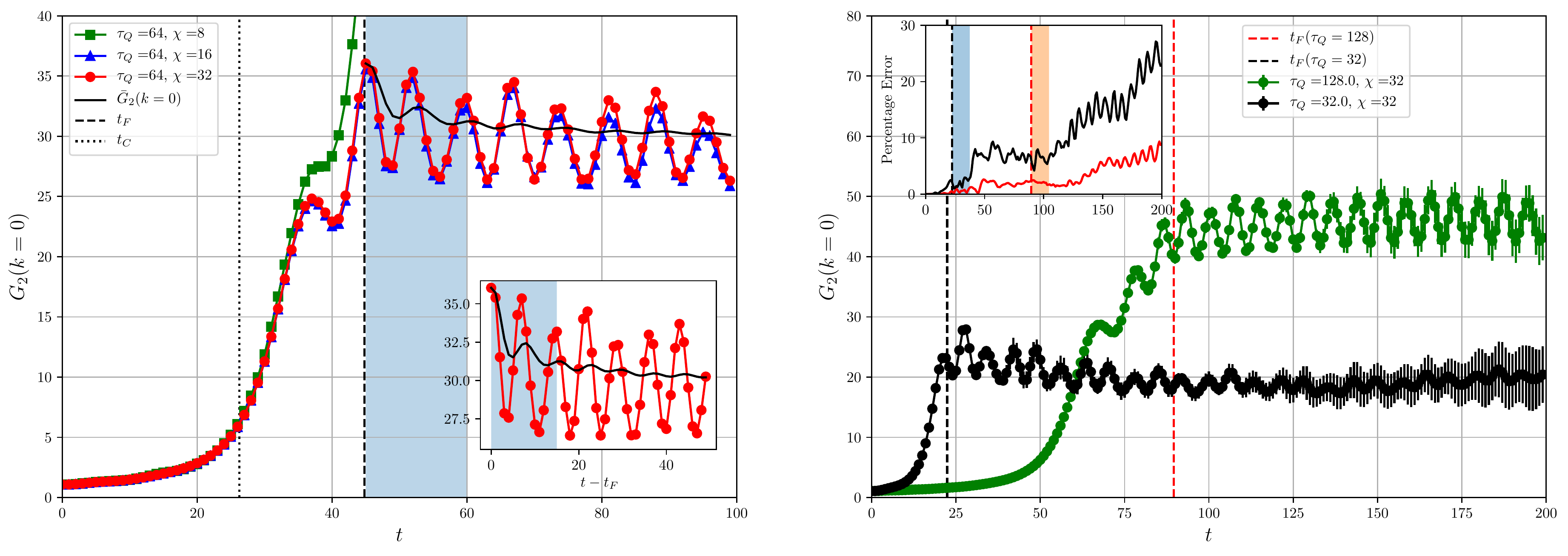}
\caption{The evolution of $G_{2}(k=0)$ is shown for $\tau_{Q} = 32, 64$ and $128$. The left-hand plot shows the approximation for $\tau_{Q} = 64$ with $\chi = 16$ (blue triangles), $\chi=32$ (red circles) and an additional $\chi = 8$ (green squares) for comparison. The $\chi = 8$ approximation deviates significantly from the $\chi = 16$ and $\chi = 32$ approximations following $t \approx t_{C}$ where $t_{C}$ is the time when $\mu_{0}^{2}(t) = m_{C}^{2}$. Until the start of relaxation at $t = t_{F}$ (vertical dashed line), the two higher $\chi$ approximations are close on this scale, but deviate visibly during the relaxation period. This is also the case for other $\tau_{Q}$ as shown in the right-hand plot where the maximum difference between the $\chi = 16,20,24,28$ and $\chi = 32$ approximations is represented by errorbars. This difference as a percentage of the value of $G_{2}(k=0)$ for $\chi = 32$ is shown in the inset where, following $t_{F}$, it can be seen that the difference becomes significant and the time-evolution has been extended up to $t=200$ to illustrate the increase of errors with time. During the relaxation period, the value of $G_{2}(k=0)$ displays large oscillations which can be removed by time averaging as shown in the left-hand plot (solid black line). While the plots show the time-evolution up to a maximum $t = 100$ for $\tau_{Q} = 64$ and $t = 200$ for $\tau_{Q} = 32, 128$, we will only be interested in comparing different $\tau_{Q}$ at the same ``relaxation time" i.e. at the same time after $t_{F}$. The maximum time used for analysis is then different for each $\tau_{Q}$ and the shaded regions indicate the data used in subsequent analysis from $t = t_{F}$ to $t = t_{F} + 15$.}
\label{EvoG20}
\end{figure*}

\begin{figure*}[t]
\centering
\includegraphics[width=1.0\linewidth]{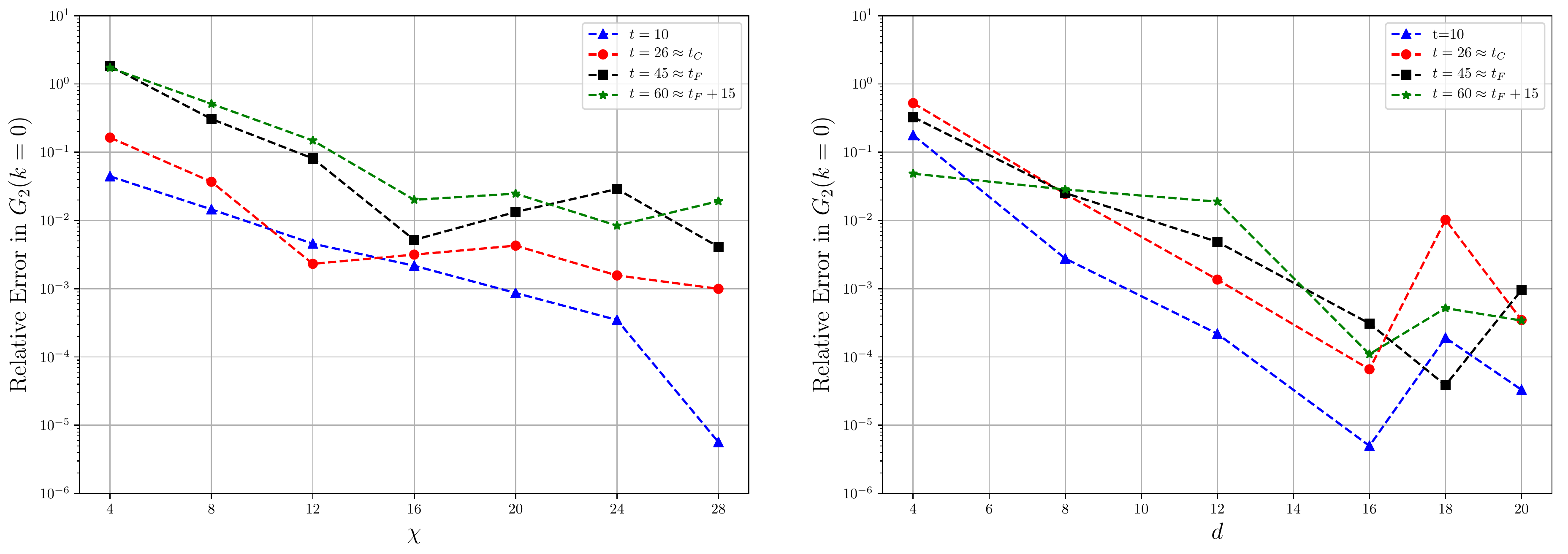}
\caption{The ``Relative Error" of $G_{2}(k=0)$, defined as the fractional difference between the value of $G_{2}(k=0)$ for a given $\chi$ or $d$ and the maximum $\chi = 32$ or $d=24$, is shown for $\tau_{Q} = 64, \mu_{0}^{2}(t_{F}) = -1.1$ and four times $t = 10, 26, 45, 60$. In the left-hand plot, the error decreases with $\chi$ for a fixed time while later times tend to have a higher overall error. However, the convergence is not smooth and, as can be seen in the $t=45$ and $t=60$ plots at $\chi = 24$, for some values of $\chi$ the error at earlier times may appear greater than at some later times. This leads to a somewhat noisy error estimate for $\chi = 32$ over time, defined as the maximum error for all $\chi \ge 16$, as can be seen in the right-hand inset of Figure \ref{EvoG20}. For $\tau_{Q} = 64$ the latest time used in subsequent analysis is $t = 60$. For all $t \le 60$ the fractional error remains less than $10^{-1}$ for all $\chi \ge 16$. In the right-hand plot, the relative error is shown for the truncation parameter $d$ with fixed $\chi = 16$. At the times $t \ge t_{F}$ used for subsequent analysis, the relative errors for $d \ge 16$ are all below $10^{-3}$ indicating that it is the error due to $\chi$ that is most relevant. As such, we simply fix $d = 18$ throughout and use the error on $\chi$ as our error estimate when performing fits.}
\label{ErrG20}
\end{figure*}

\section{Results}
\label{Res}

We study the non-equilibrium dynamics of the $\phi^{4}$ quantum field theory using tensor network techniques. In particular, we study the lattice regularised Hamiltonian (\ref{HamLat}) with a time-dependent bare mass (where we drop the tildes for notational convenience)
\begin{align}
\mu_{0}^{2}(t) = -\frac{t}{\tau_{Q}} + \mu_{0}^{2}(t=0) ~ , ~ t < t_{F} \nonumber \\
\mu_{0}^{2}(t) = \mu_{0}^{2}(t_{F}) ~ , ~ t \ge t_{F} ~ .
\end{align}
This time-dependence drives a ground state $\ket{\Omega(\mu_{0}^{2}(t=0))}$ from the symmetric phase $\mu_{0}^{2}(t=0) > m_{C}^{2}$ into the broken symmetry phase $\mu_{0}^{2}(t_{F}) < m_{C}^{2}$ where it relaxes under a time-independent Hamiltonian with bare mass $\mu_{0}^{2}(t_{F})$.  The initial ground-state is approximated by a uMPS using the VUMPS algorithm while the time-evolution is approximated by evolving the initial uMPS according to the TDVP projected Schr\"{o}dinger equation (\ref{SE_UMPS}) using a $5^{\text{th}}$ order Runge-Kutta scheme.

The physics of this non-equilibrium scenario is described by the Kibble-Zurek mechanism of topological defect formation and we compare the uMPS approximation of the equal-time two-point function $G_{2}(k) = \braket{\psi(t)| \phi(-k) \phi(k) |\psi(t)}$, obtained via a discrete cosine transform of $G_{2}(r)$, to the KZM expectations. In particular, we assume that the non-equilibrium two point function is approximated by the defect ansatz (\ref{DefectAnsatz}) such that $G_{2}(k) \approx G_{\text{def}}(k)$ and we check the consistency of this assumption in several stages. Firstly, we compare the initial evolution of $G_{2}(k=0)$ within the symmetric phase, which provides a measure of the correlation length $\xi$, to its equilibrium value $G_{2}^{\Omega}(k=0)$. We confirm that equilibrium is lost at a distance $\hat{\epsilon} = \hat{\mu}_{0}^{2} - m_{C}^{2}$ from the critical point and that $\hat{\epsilon}$ scales as expected with the quench rate $\tau_{Q}$. Secondly, we study the time averaged equal time two point function $\bar{G}_{2}(k=0)$ in the broken symmetry phase, confirming that it also scales as expected and provides a consistent estimate of the defect density $n_{\text{est}}$ (\ref{nEst_Def}) under the assumption that $\bar{G}_{2}(k=0) = G_{\text{def}}(k=0)$. Thirdly, we show that the observable $G_{\text{uni}}(k)$ (\ref{GuniQ}) is a universal function of $n$ for low $k$ and that it is described by the functional form of $G_{\text{corr}}(k/n)$ (\ref{Def_uniK}). Furthermore, by including the contribution of the kink profile via the semi-classical approximation of $G_{\text{kink}}(k d_{K})$, the function $G_{\text{uni}}(k)/G_{\text{kink}}(k d_{K})$ is also a universal function of $n$ but now over a larger region of $k$. Finally, we show that $\bar{G}_{2}(k) \approx G_{\text{def}}(k)$ for all $k$ by including the matter contributions $G_{\text{mat}}(k)$ (\ref{ThermAns}) to the defect ansatz via a two-parameter fit.

To justify the explicit set up used to study the KZM (i.e. the choices of $\mu_{0}^{2}(0), \mu_{0}^{2}(t_{F})$ and $\tau_{Q}$) we can examine the equilibrium physics of the theory. The important parameter regions can be identified by fixing the bare coupling and producing a series of uMPS approximations to the ground-state $\ket{\Omega[A]}$. The scalar mass $m_{S}$ can then be extracted from these approximations using the one-particle excitation ansatz described in \cite{Vanderstraeten2015}. The kink mass $M_{K}$ can also be obtained via a similar ansatz \cite{Haegeman2012}, though here we use the PBC MPS method described in \cite{Gillman2017}. These approximations to the scalar mass and kink mass are plotted in Figure \ref{EquilMS} which demonstrates the various regions of interest.

\begin{figure*}[t]
\centering
\includegraphics[width=1.0\linewidth]{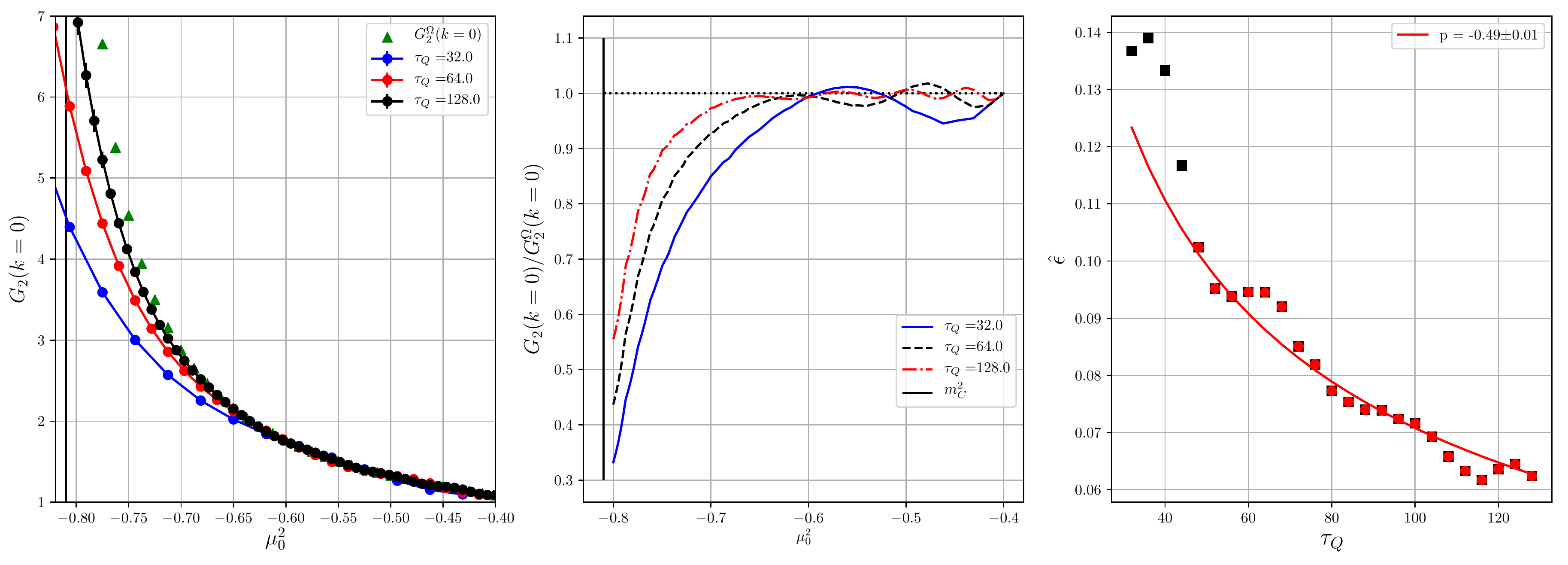}
\caption{Plots of the time evolution of $G_{2}(k=0)$ up the critical point $m_{C}^{2}$ (vertical solid black line). In the leftmost plot this is compared with the ground state value (green triangles) with the increasing $\tau_{Q}$ remaining close to this value for a larger region of $\mu_{0}^{2}$. In the central plot, the ratio of $G_{2}(k=0)$ to the ground-state value is plotted using interpolating functions, showing the departure from equilibrium more clearly.  This also allows for a criteria for the loss of equilibrium to be established and we used the condition $G_{2}(k=0)/G_{2}^{\Omega}(k=0) = 0.9$ to define $\hat{\epsilon} = \hat{\mu}_{0}^{2} - m_{C}^{2}$ as the point where equilibrium is lost. The value of $\hat{\epsilon}$ is shown in the rightmost plot where the larger $\tau_{Q}$ data (red circles) are fit to a power-law.}
\label{AdiG2}
\end{figure*}

\begin{figure*}[t]
\centering
\includegraphics[width=1.0\linewidth]{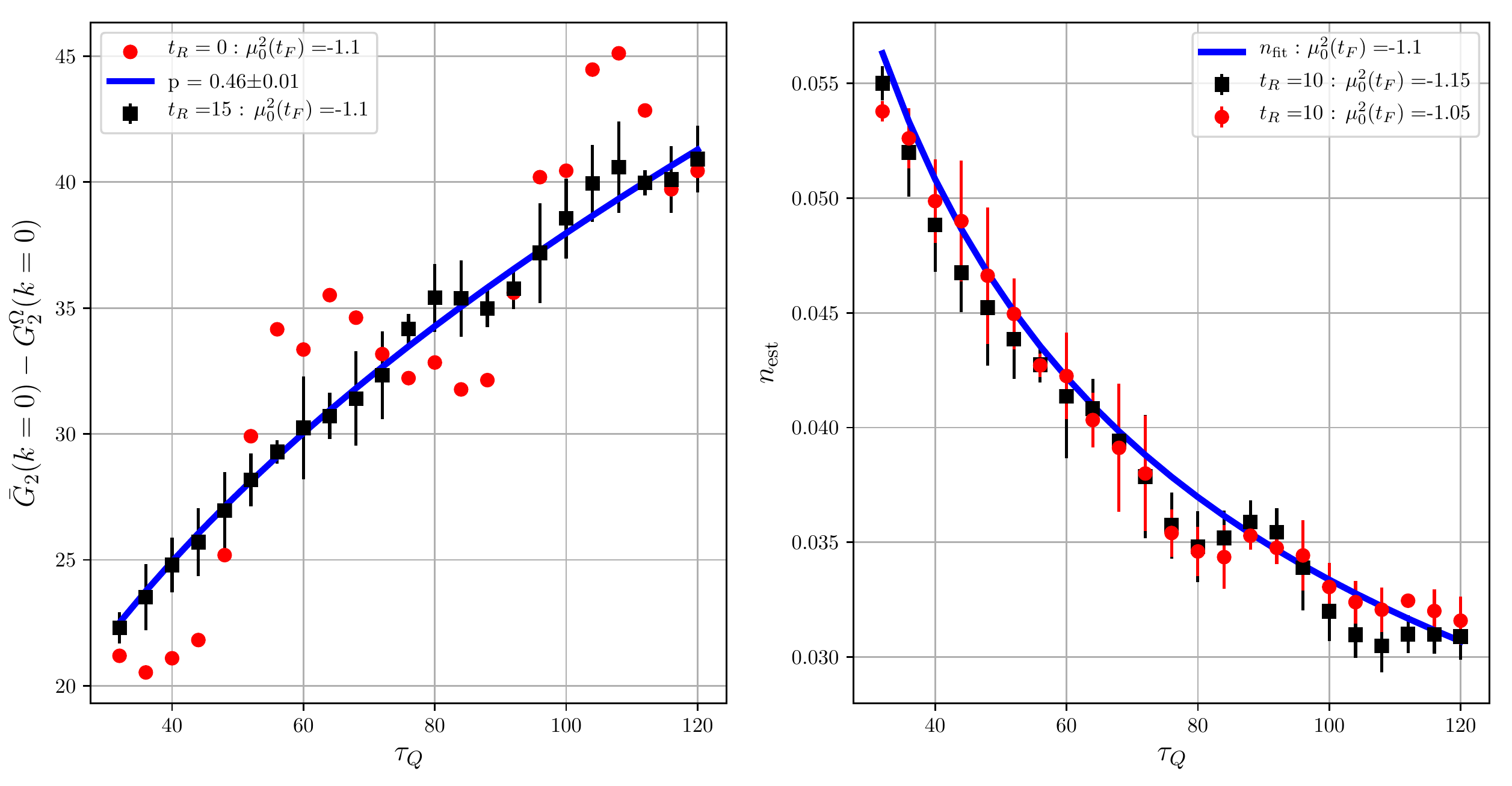}
\caption{The value of $\bar{G}_{2}(k=0)$ with the vacuum subtracted is shown (left-hand plot) for $t_{R} = 0$ and $t_{R}=15$. In the former case, there are large oscillations present in the data as different $\tau_{Q}$ quenches lie in different phases of their evolution. These oscillations are significantly damped at $t_{R}=15$  by the time-averaging and a power-law fit has been taken. The defect density corresponding to this power-law fit $n_{\text{fit}}$ is shown in the right-hand plot which can be compared to the estimates $n_{\text{est}}$ (\ref{nEst_Def}) for $\mu_{0}^{2}(t_{F})=-1.05,-1.15$ at $t_{R}=10$. The data lies fairly close to the fit though the larger $\tau_{Q}$ data still displays clear oscillations.}
\label{Nest}
\end{figure*}

To approximate the behaviour of the quantum field theory in the lattice regularised setting, we will be interested in working in the ``continuum region" such that $\xi > 1$ corresponding to $m_{S} < 1$. In this region, the effects of the lattice regularisation will be small and we will only consider evolutions that take place within this region. Furthermore, we are interested in setting $\mu_{0}^{2}(t_{F})$ to lie outside the ``strong coupling " region in the broken symmetry phase where $m_{S} = 2 M_{K}$ and kink-antikink pairs do not behave like classical extended objects but as standard particles. These two considerations then both limit the potential choices of $\mu_{0}^{2}(t_{F})$ and we have indicated the set of $\mu_{0}^{2}(t_{F})$ we use in Figure \ref{EquilMS}. The initial $\mu_{0}^{2}(0)$ is also chosen to lie outside the strong-coupling region in the symmetric phase and a set of quench rates $\tau_{Q}$ are chosen so that it is possible to maintain equilibrium into the strong-coupling region such that the scaling arguments from the KZM can be applied.  While increasing the bare coupling $\lambda_{0}$ enlarges the strong-coupling region so that lower $\tau_{Q}$ are required, it also shrinks the available continuum region in the broken symmetry phase that lies outside the strong-coupling region, and we have found $\lambda_{0} = 3$ to provide a good balance. 

\begin{figure*}[t]
\centering
\includegraphics[width=1.0\linewidth]{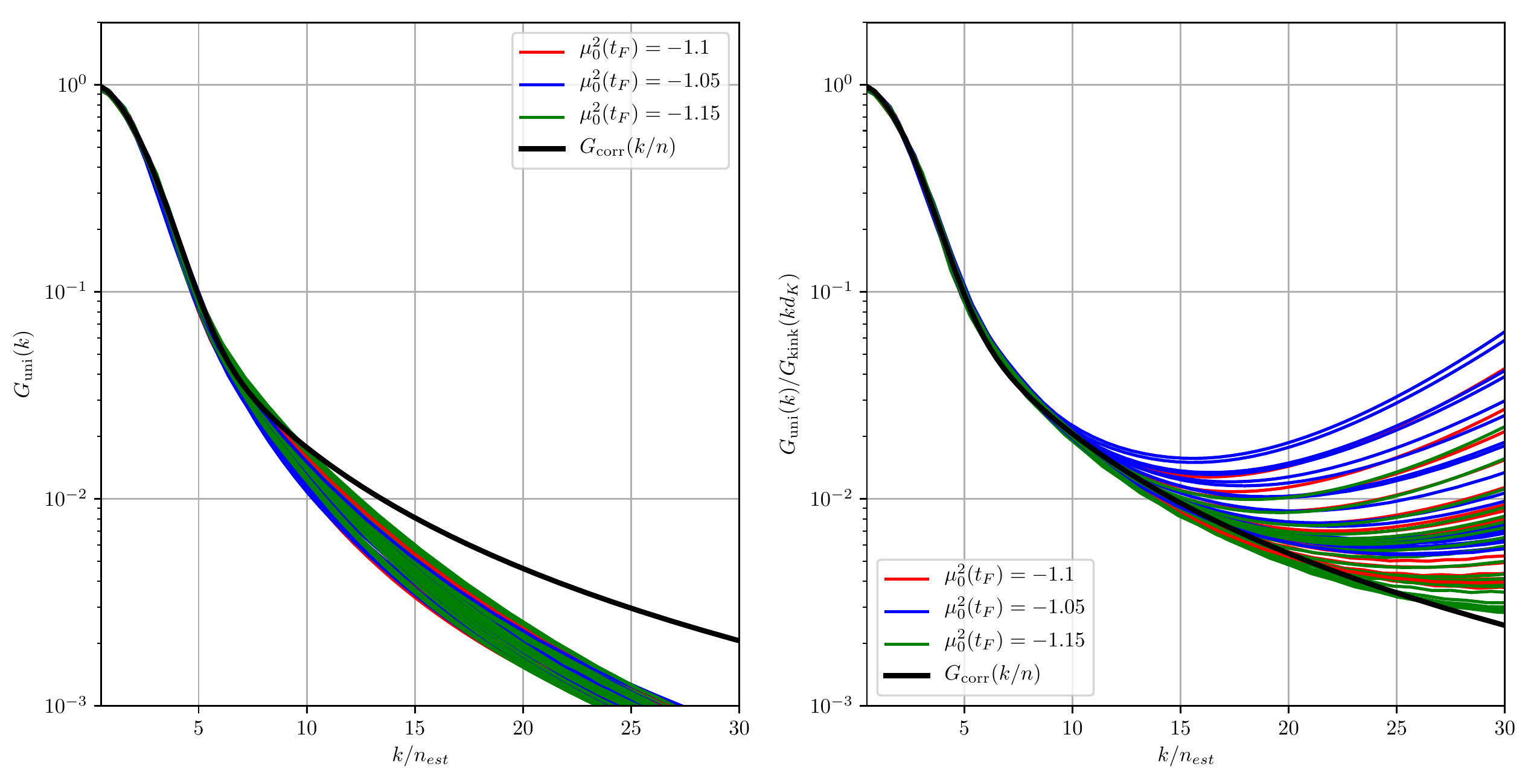}
\caption{Plots of $G_{\text{uni}}(k)$ (\ref{GuniQ}) (left-hand plot) and  $G_{\text{uni}}(k)/G_{\text{kink}}(k d_{K})$ (right-hand plot) for $\tau_{Q} = 32,36,40,...,128$ along with fits to the form functional form $G_{\text{corr}}(k)$ (\ref{ClassicalFit}). In the left-hand plot, the data collapses up to $k/n_{\text{est}} \approx 5$ indicating a universal function of the defect density up to this point. With the inclusion of kink profile $G_{\text{kink}}(k d_{K})$ in the right-hand plot the universal region is increased to $k/n_{\text{est}} \approx 10$ indicating that the semi-classical approximation for $G_{\text{kink}}(k d_{K})$ is accurate and that the data is consistent with kink formation in the system. The fitted form agrees well with the data in both cases within the universal regions giving $\alpha_{1},\alpha_{2},\beta_{1},\beta_{2} = 0.683,0.120,0.329,0.176$ and $\alpha_{1},\alpha_{2},\beta_{1},\beta_{2} = 0.723,0.128,0.290,0.130$ in the left-hand and right-hand plots respectively.}
\label{GuniFig}
\end{figure*}

With the parameters for the evolutions fixed and an approximation of the initial state $\ket{\Omega[A]_{0}}$ obtained by uMPS with bond-dimension $\chi$, the equal-time two function can be approximated by numerically integrating the evolution equation (\ref{SE_TenUpd}). We do this for a set of $\chi = 16,20,24,28,32$ using a $5^{\text{th}}$ order Runge-Kutta scheme with time step $\tau = 10^{-2}$ and local basis truncation $d=18$ up to total time $T = 100$ with the observable $G_{2}(k)$ being evaluated every $100$ steps. The time evolution of $G_{2}(k=0)$ for the case of $\mu_{0}^{2}(t_{F}) = -1.1$ and $\tau_{Q} = 32,64,128$ is shown in Figure \ref{EvoG20} which illustrates several features of the evolution and approximations used. 

In the left-hand plot of Figure \ref{EvoG20}, the value of $G_{2}(k=0)$ is shown for $\tau_{Q} = 64$ and $\chi = 16 , 32$. Initially, the difference between the two is small, being almost indistinguishable on this scale prior to the critical point (dotted vertical line) but the difference becomes significant during the ``relaxation" portion of the evolution at $t > t_{F}$ (dashed vertical line). Ideally, we would like to make a set of approximations for different $\chi$ and extrapolate to the $\chi \to \infty$ limit where the evolution of the regularised theory is exact. While this is possible in some cases where the convergence of observables is particularly smooth, it is difficult in others especially at later times when we no longer expect the state itself to be well described by uMPS with limited bond-dimension, even if the observable of interest itself can be reasonably approximated. As such, we instead make a simple estimate of the error by taking the maximum absolute difference between the $\chi = 16,20,24,28$ and highest $\chi = 32$ approximations which we use as an input when fitting curves and display as errorbars in plots (see Figure \ref{ErrG20} for more discussion of the errors due to $\chi$ and $d$). The right-hand plot of Figure \ref{EvoG20} shows the evolution of the two $\tau_{Q} = 32$ and $\tau_{Q} = 128$ quenches along with the errorbars. Once again, the most significant errors occur at later times $t \gg t_{F}$ as the system relaxes. In the $\tau_{Q} = 32$ cases, there are then significant errors occurring at much earlier absolute time $t$ than for the corresponding $\tau_{Q} = 128$, as shown in the inset which gives the error as a percentage of the value of $G_{2}(k=0)$. However, we will not be interested in comparing different $\tau_{Q}$ quenches at the same absolute time but rather at the same relaxation time $t_{R} = t - t_{F}$ after the quench ends. As such we will not be interested in the regions with the most significant errors far from the point $t_{F}$ (the dashed black and red vertical lines) such that the errors in different $\tau_{Q}$ quenches will be much closer than if taken at the same absolute time. 

In addition to the errors, the plots in Figure \ref{EvoG20} display temporal oscillations in $G_{2}(k=0)$, particularly during the relaxation period, that are characteristic of the non-equilibrium dynamics of quenched systems, as often found when studying instantaneous $\tau_{Q} \to 0$ quenches \cite{Sotiriadis2010}. While there may be some physical damping of these oscillations over time, the time-scale on which this occurs is longer than the time-scales we have approximated. Because of this, rather than focus on the equal-time two point function directly, we will instead use the time-averaged two point function $\bar{G}_{2}(k)$ given by averaging over the available data for $G_{2}(k)$ after the final bare mass $\mu_{0}^{2}(t_{F})$ has been reached and relaxation begins such that
\begin{align}
\overline{G_{2}}(k,t_{R}) = \frac{1}{t_{R}} \int_{t_{F}}^{t_{F}+t_{R}} G_{2}(k,T) ~ dT . 
\label{G2_Avg}
\end{align}
This observable can then be used to provide a clean comparison with the expected KZM behaviour and displays a similar error for different $\tau_{Q}$ quenches given a fixed relaxation time $t_{R}$. The value of $\bar{G}_{2}(k=0)$ for the $\tau_{Q} = 64$ case with $\chi = 32$ is shown in the left-hand plot of Figure 2 with the inset displaying some of the region $t > t_{F}$ where the averaging can be clearly seen. 

\begin{figure*}[t]
\centering
\includegraphics[width=1.0\linewidth]{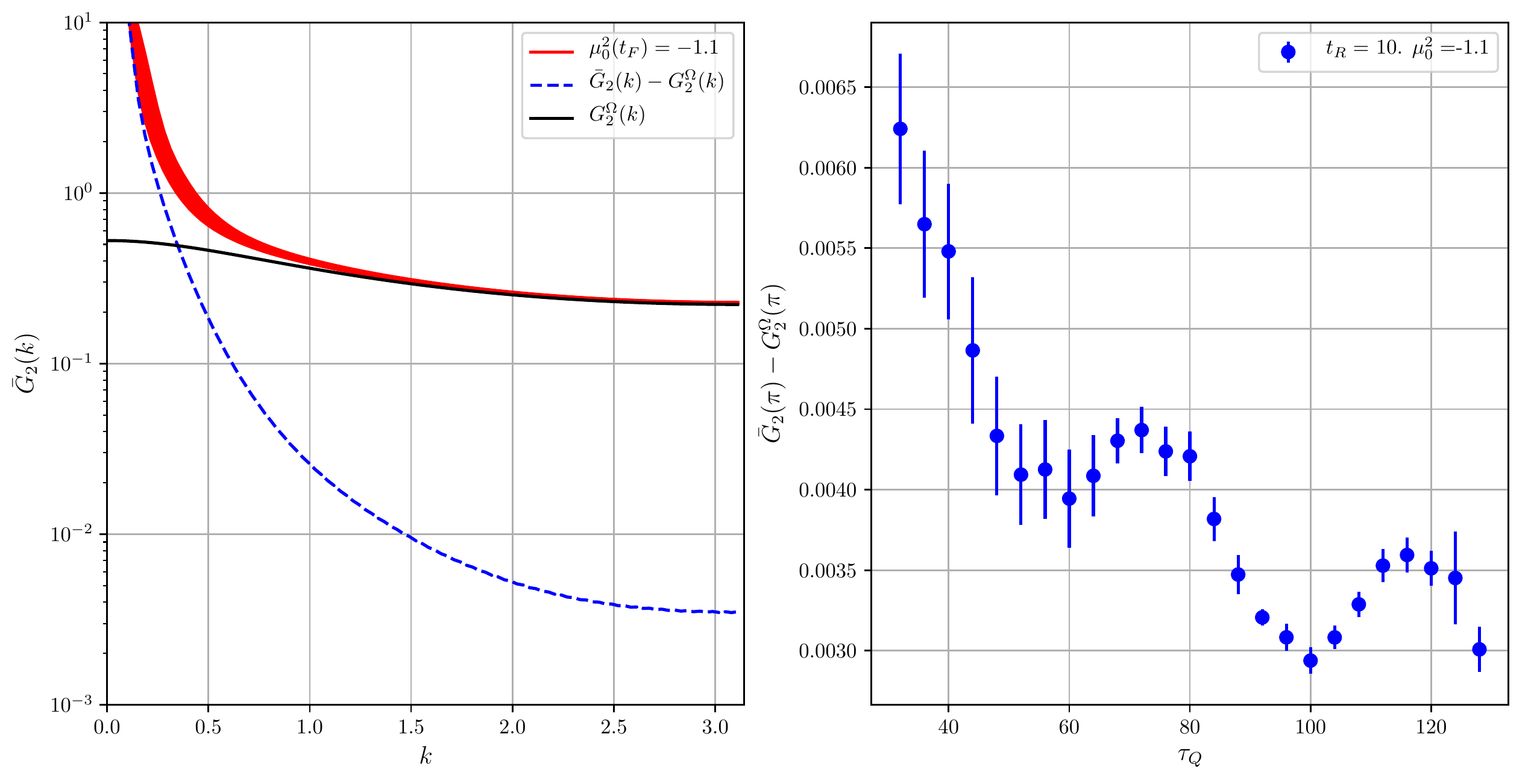}
\caption{The value of $\bar{G}_{2}(k)$ for $\tau_{Q} = 32,36,40,...,128$ (solid red lines) is plotted along with $G_{2}^{\Omega}(k)$ (left-hand plot, solid black line) which agrees closely at large $k$ as illustrated by the difference $\bar{G}_{2}(k) - G_{2}^{\Omega}(k)$ (dashed blue lines). At the maximum momentum $k=\pi$ the value of $\bar{G}_{2}(k=\pi)$ still lies above the vacuum value but this positive contribution decreases with $\tau_{Q}$ (right-hand plot, blue circles) consistent with the existence of additional non-vacuum contributions to $\bar{G}_{2}(k)$ that are suppressed  by slower quench rates.}
\label{G2_Vac}
\end{figure*}

\begin{figure*}[t]
\centering
\includegraphics[width=1.0\linewidth]{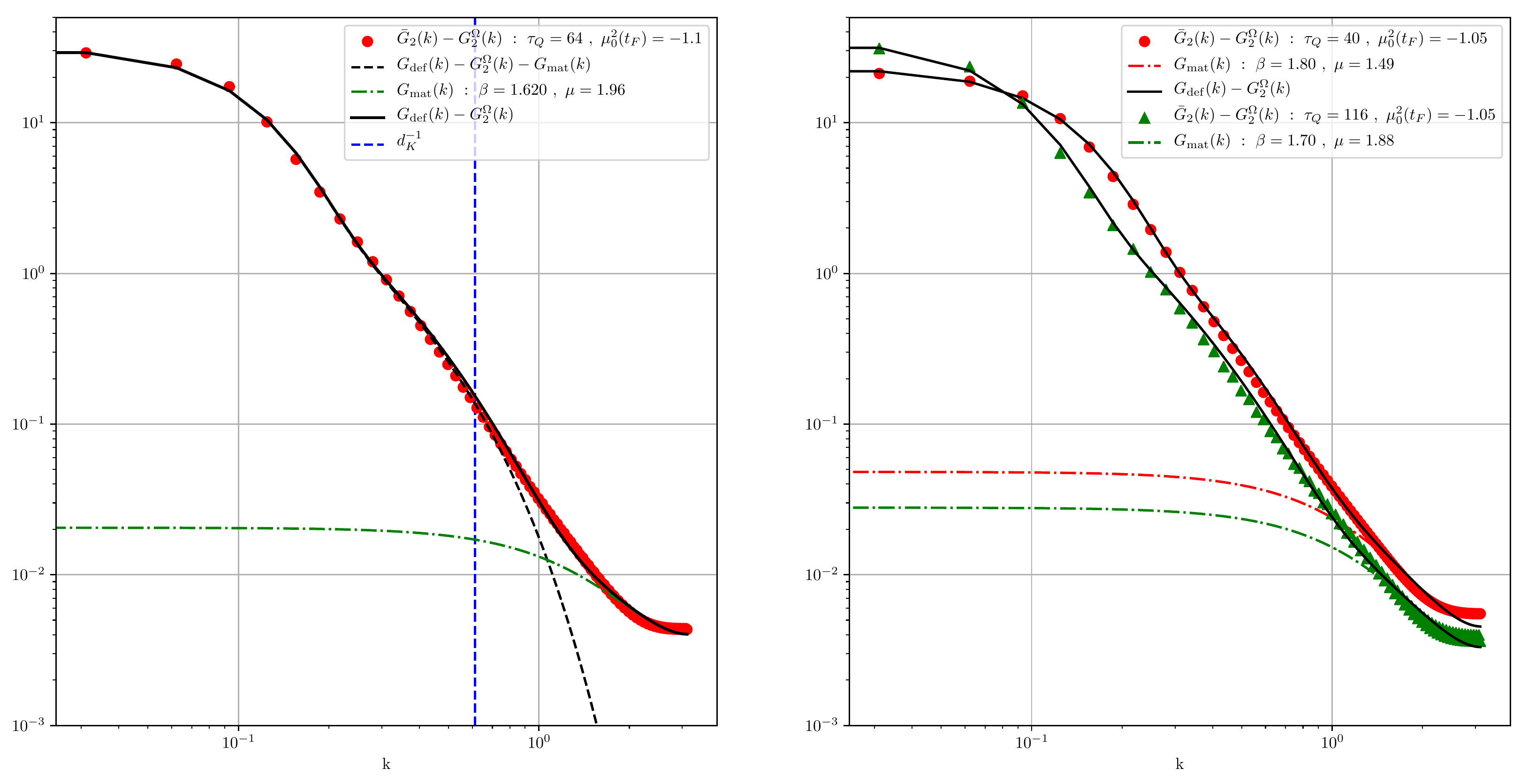}
\caption{The value of $\bar{G}_{2}(k) - G_{2}^{\Omega}(k)$ for $\tau_{Q} = 64 , \mu_{0}^{2}(t_{F}) = -1.1$ (left-hand plot, red circles) and $\tau_{Q} = 40 , 116,~ \mu_{0}^{2}(t_{F}) = -1.05$ (right-hand plot, red circles and green triangles respectively). All three observables are fitted to $G_{\text{def}}(k) - G_{2}^{\Omega}(k)$ (\ref{DefectAnsatz}) via a two parameter fit with the values of $\beta , \mu$ indicated. The $G_{\text{mat}}(k)$ (\ref{ThermAns}) component of $G_{\text{def}}(k) - G_{2}^{\Omega}(k)$, which contains the free parameters, is shown in both plots (dotted-dashed line). Initially, this component is irrelevant and the curves are described by a single universal part shown in the left-hand plot (black dashed line). In the universal part the effect of finite width kinks is clearly seen with the form being exponentially suppressed after the inverse kink width $k \approx d_{K}^{-1}$ (vertical blue dashed line), such that the matter component dominates in the higher $k$ region. In the right-hand plot the curves behave as expected with the higher $\tau_{Q}$ case having a smaller matter component.}
\label{G2_Therm}
\end{figure*}

To check the initial evolution of the system up to the critical point, we compare the behaviour of $G_{2}(k=0)$ to the equilibrium value $G_{2}^{\Omega}(k=0)$. The KZM states that initially the state should remain in equilibrium such that $G_{2}(k=0) = G_{2}^{\Omega}(k=0)$ before becoming excited at some point $\hat{\epsilon} = \mu_{0}^{2} - m_{C}^{2}$ before the critical point. The KZM further provides an estimate for the scaling of $\hat{\epsilon}$ with the quench rate $\tau_{Q}$ such that $\hat{\epsilon} \sim \tau_{Q}^{-1/2}$ when $\tau_{Q}$ is sufficiently large to probe the critical region. Figure \ref{AdiG2} illustrates this behaviour through the evolution of $G_{2}(k=0)$ (leftmost plot), the ratio $G_{2}(k=0)/G_{2}^{\Omega}(k=0)$ constructed using interpolating functions (centre plot) and a comparison of $\hat{\epsilon}$ with $\tau_{Q}$, where $\hat{\epsilon}$ is estimated by the point at which $G_{2}(k=0)/G_{2}^{\Omega}(k=0) = 0.9$ (rightmost plot). The scaling of $\hat{\epsilon}$ is established using a power-law fit to the points shown in red to give $\hat{\epsilon} \sim \tau_{Q}^{-0.49 \pm 0.01}$ close to the $\tau_{Q}^{-1/2}$ predicted by the KZM.

As the system enters the broken symmetry phase, the value of $G_{2}(k=0)$ continues to grow, but the scaling established at the point $\hat{\epsilon}$ should be retained. Furthermore, once in the symmetry broken phase and after sufficient relaxation time, we can interpret the time-average $\bar{G}_{2}(k=0)$ via the defect ansatz $G_{\text{def}}(k=0)$ such that $\bar{G}_{2}(k=0) - G_{2}^{\Omega}(k=0) \approx v^{2}/n$, where the vacuum expectation value $v(\mu_{0}^{2})$ is determined from the corresponding uMPS approximation of the ground-state $v = \braket{\Omega[A]| \phi | \Omega[A]}$.  The value of $\bar{G}_{2}(k=0)  - G_{2}^{\Omega}(k=0)$ is shown in Figure \ref{Nest} (left-hand plot) for the $\mu_{0}^{2}(t_{F}) =-1.1$ case with $t_{R} = 0$ and $t_{R} = 15$. In the first case, the oscillatory behaviour is clearly visible as different $\tau_{Q}$ lie at different phases of their evolution. However, in the second case this behaviour is damped significantly by the time-averaging. A power-law fit of the $t_{R} = 15$ data scales as $\tau_{Q}^{0.46 \pm 0.01}$. This value is somewhere between the classical $\tau_{Q}^{1/3}$ and quantum $\tau_{Q}^{1/2}$, though closer to the latter. The estimate of the defect density $n_{\text{est}}$ (\ref{nEst_Def}) is shown in the right-hand plot of Figure \ref{Nest} for the cases $\mu_{0}^{2}(t_{F}) = -1.15$ and $-1.05$. The agreement between the different $\mu_{0}^{2}(t_{F})$ and $n_{\text{est}}$ obtained from the power-law fit of $\bar{G}_{2}(k=0) - G_{2}^{\Omega}(k=0)$ is reasonable, consistent with the interpretation that $\bar{G}_{2}(k=0) \approx v^{2}/n$, though the large $\tau_{Q}$ data still displays oscillations for this $t_{R}$ and the low $\tau_{Q}$ data for $\mu_{0}^{2}(t_{F}) = -1.15$ lies somewhat below $n_{\text{fit}}$ indicating that longer relaxation times and slower quenches could improve the agreement further. Nevertheless, this data suggests that $\bar{G}_{2}(k=0)$ can indeed provide a simple observable with which to estimate the defect density in a quantum field theory. 

With the defect density estimated, the comparison of $\bar{G}_{2}(k)$ and $G_{\text{def}}(k)$ can continue by scaling the data using the estimated $n_{\text{est}}$ and examining the observables $G_{\text{uni}}(k)$ (\ref{GuniQ}) and $G_{\text{uni}}(k)/G_{\text{kink}}(k d_{K})$ using the semi-classical approximation for $G_{\text{kink}}(k d_{K})$ described in Section \ref{KZM}. If indeed $\bar{G}_{2}(k) = G_{\text{def}}(k)$ then $G_{\text{uni}}(k)$ should be a universal function of $n$ up to a scale where the defect width $d_{K}$ is important and $G_{\text{uni}}(k) \approx G_{\text{corr}}(k/n)$. If additionally the approximate form of $G_{\text{kink}}(k d_{K})$ is accurate, then we can further expect the observable $G_{\text{uni}}(k)/G_{\text{kink}}(k d_{K})$ to be a universal function of $n$ up to a somewhat higher scale where the matter contributions $G_{\text{mat}}(k)$ become important and $G_{\text{uni}}(k)/G_{\text{kink}}(k d_{K}) \approx G_{\text{corr}}(k/n)$ over this region.

Figure \ref{GuniFig} shows the two observables $G_{\text{uni}}(k)$ (left-hand plot) and $G_{\text{uni}}(k)/G_{\text{kink}}(k d_{K})$ (right-hand plot) for $\tau_{Q} = 32,36,40,...,128$ and $\mu_{0}^{2}(t_{F}) = -1.05,-1.1,-1.15$ along with fits to the functional form of $G_{\text{corr}}(k)$. In the left-hand plot of Figure \ref{GuniFig} the observable $G_{\text{uni}}(k)$ collapses reasonably up to around $k/n_{\text{est}} \approx 5$ and the functional form of $G_{\text{corr}}(k)$ fits well in this region such that the approximation
\begin{align}
\bar{G}_{2}(k) \approx \frac{v^{2}}{n_{\text{est}}}G_{\text{corr}}(k/n_{\text{est}}) + G_{2}^{\Omega}(k) ~
\end{align}
holds for these low $k/n_{\text{est}}$. However, for larger $k/n_{\text{est}}$ the data begins to spread out indicating that $G_{\text{uni}}(k)$ is not a universal function of $n$ in this region. Furthermore, the fit sits above the data indicating the need for an additional term to suppress it and suggesting that there is another relevant scale missing.

According to the Kibble-Zurek mechanism and the physical picture provided by the defect ansatz (\ref{DefectAnsatz}), this missing scale should be given by the width of defects in the system $d_{K}$. In the right-hand plot of Figure \ref{GuniFig}, the observable $G_{\text{uni}}(k)/G_{\text{kink}}(k d_{K})$ collapses well up until $k/n_{\text{est}} \approx 10$. At this point the data spreads out and begins to increase due to the division of $G_{\text{kink}}(k d_{K})$ which becomes small in this region. Nevertheless, the fit still agrees at $k/n_{\text{est}} \approx 20$ with a number of curves for which the division by $G_{\text{kink}}(k d_{K})$ has not yet dominated. Up to this scale, we then have the approximation that
\begin{align}
\bar{G}_{2}(k) \approx \frac{v^{2}}{n_{\text{est}}}G_{\text{corr}}(k/n_{\text{est}})G_{\text{kink}}(k d_{K}) + G_{2}^{\Omega}(k) ~
\label{NoMat}
\end{align}
which is the defect ansatz (\ref{DefectAnsatz}) with the matter contribution $G_{\text{mat}}(k)$ neglected.

To further check the consistency of the approximation $\bar{G}_{2}(k) \approx G_{\text{def}}(k)$ we would like to account for the matter contributions. Firstly, we can check the consistency of their interpretation by comparing the data $\bar{G}_{2}(k)$ to the vacuum explicitly as shown in Figure \ref{G2_Vac}. As expected, the equal time two point function tends to the vacuum at high $k$ for all $\tau_{Q}$ but has an additional positive contribution that is suppressed with increasing $\tau_{Q}$, consistent with the generation of additional non-vacuum excitations during the phase transition which provide the contribution $G_{\text{mat}}(k)$ to the equal time two point function.

We can now account for the remaining contributions to $\bar{G}_{2}(k)$ by using a semi-classical ansatz for the matter contributions $G_{\text{mat}}(k)$ (\ref{ThermAns}) as discussed in Section \ref{KZM}. This constitutes a two parameter fit and we find that, once performed, the approximation $\bar{G}_{2}(k) \approx G_{\text{def}}(k)$ holds over several orders of magnitude, as shown in Figure \ref{G2_Therm}.

Figure \ref{G2_Therm} displays the defect ansatz fit (\ref{DefectAnsatz}) (solid black line) with the vacuum subtracted for the $\tau_{Q} = 64, \mu_{0}^{2}(t_{F}) = -1.1$ data (red circles) along with the various components of the fit (left-hand plot). Firstly, the defect ansatz without the matter component (\ref{NoMat}) (dashed black line) decays rapidly to zero following the scale set by the kink width $d_{K}^{-1} \approx 0.61$ (vertical dashed blue line). This is corrected by the matter contribution shown (dotted-dashed line) which is initially irrelevant but dominates at high $k \gg d_{K}^{-1}$. The full defect ansatz (solid black line) then approximates the data reasonably over the full range of $k$. The right-hand plot also shows the fits for the case $\mu_{0}^{2} = -1.05$ with lower $\tau_{Q} = 40$ and higher $\tau_{Q} = 116$ data (red circles and green triangles respectively). In this case, the plots behave as expected with the higher $\tau_{Q}$ data starting at a larger value for low $k$ , corresponding to a lower defect density, but ending up at a lower value since there are fewer non-vacuum excitations present.

While the defect ansatz provides a reasonable approximation with the form of $G_{\text{mat}}(k)$ given by $(\ref{ThermAns})$, $(\ref{OmegaAns})$ and the corresponding values of $\mu$ and $\beta$ determined by fitting, we cannot interpret these parameters as cleanly as we would like since they display large variations with $\tau_{Q}$ that mask any overall trend. This also somewhat obscures the interpretation of the kink profile term since we cannot assess the impact of the semi-classical approximation cleanly. To improve this, it would be desirable in the future to have a non-perturbative approximation for the matter contribution $G_{\text{mat}}(k)$. This can be achieved by assuming, as done in this paper, that the matter contributions can be described by thermal effects. The thermal two-point function can then be estimated by a non-perturbative method such as the \textit{minimally entangled thermal states} (METTS) tensor network \cite{Stoudenmire2010}, which takes $\beta$ as an input with a definite interpretation as the inverse temperature and eliminates the need for the additional parameter $\mu$. While it is still not clear exactly what value of $\beta$ should be used since we do not know how energy is partitioned in the system, this would still offer a more rigorous result and we could, e.g. determine $\beta$ by fitting to the data at large $k$ where any effect of the kink profile should be irrelevant. If we are able to establish the form of $G_{\text{mat}}(k)$ in this manner, we can then ``measure" $G_{\text{kink}}(k d_{K})$ more directly. In principle, this can then be compared with a non-perturbative approximation of  $G_{\text{kink}}(k d_{K})$ which e.g. might be obtained though a TN approximation of the equal time two point function of the one kink state $\braket{K|\phi(-k)\phi(k)|K}$. 

\section{Conclusion}
\label{Concl}

We have studied the relativistic $\phi^{4}$ scalar field theory in $D=(1+1)$ as the system is driven through a quantum phase transition and approximated the equal time momentum space two point function $G_{2}(k)$ using uniform matrix product states. 

We have compared the time averaged two point function within the symmetry broken phase $\bar{G}_{2}(k)$ to the ansatz $G_{\text{def}}(k)$ (\ref{DefectAnsatz}) based on the expectation of universal formation of kinks via the Kibble Zurek mechanism. We find that $\bar{G}_{2}(k)$ contains a universal part $G_{\text{uni}}(k)$ that is a universal function of the estimate of the defect density $n_{\text{est}}$ (\ref{nEst_Def}) for low $k$ and that the functional form agrees with that of $G_{\text{def}}(k)$ using a semi-classical approximation for the kink profile $G_{\text{kink}}(k)$. The approximation $\bar{G}_{2}(k) \approx G_{\text{def}}(k)$ further holds reasonably for all $k$ with the inclusion of a semi-classical ansatz $G_{\text{mat}}(k)$ for the matter contributions to $G_{\text{def}}(k)$ which then constitutes a two parameter fit. These results indicate the consistency of the picture that the state of the system following a symmetry breaking phase transition is indeed described by the Kibble Zurek mechanism along with the defect ansatz (\ref{DefectAnsatz}) and that tensor network techniques can capture the non-perturbative non-equilibrium physics of topological defect formation in relativistic quantum field theories with strong-coupling quantum phase transitions.

While the approximations used mean that precise quantitative predictions are challenging, overall, when considering the difficulty of performing calculations of topological defect formation in quantum field theory, our results suggests that the future application of tensor network techniques to the study of non-perturbative, non-equilibrium effects in QFT is highly promising. We have suggested some possible improvements that will allow for more quantitative predictions in the future and, with the rapid recent developments of tensor network techniques, are hopeful that these techniques can be applied in the near future to more realistic models within high energy physics and cosmology, with non-perturbative non-equilibrium scenarios being an area where they can offer particular advantage over other available techniques.

\section*{Acknowledgements}

We have made use of the Imperial College London High Performance Computing Service. E.G. was supported by the EPSRC Centre for Doctoral Training in Controlled Quantum Dynamics and A.R. by STFC grant ST/L00044X/1.

\bibliography{TD_KZM_MPS.bib}

\end{document}